\documentclass[
amssymb,preprint,preprintnumbers,nofootinbib,superscriptaddress]{revtex4-2}
\usepackage{amsmath}
\usepackage{graphicx}
\graphicspath{{Figures/}}
\usepackage{latexsym}
\usepackage{amsfonts}
\usepackage{url,hyperref}
\usepackage{bm}
\usepackage{textcomp}
\usepackage{xcolor}
\usepackage{bbm}
\usepackage{slashed}
\usepackage{caption}
\usepackage{epstopdf}
\usepackage{subcaption}
\usepackage{placeins}
\usepackage{color}
\usepackage[a4paper,left=20mm,right=20mm,top=25mm,bottom=25mm]{geometry}

\begin{document}
\title{Thermodynamics of Black Holes  with R\'enyi Entropy from Classical Gravity}

\author{Ratchaphat Nakarachinda 
\footnote{Email: tahpahctar\_net@hotmail.com}}
\affiliation{Department of Mathematics and Computer Science, Faculty of Science, Chulalongkorn University, Bangkok 10330, Thailand}	
\affiliation{The Institute for Fundamental Study (IF), Naresuan University,\\ 99 Moo 9, Tah Poe, Mueang Phitsanulok, Phitsanulok, 65000, Thailand}

\author{Chatchai Promsiri
\footnote{Email: chatchaipromsiri@gmail.com}}
\affiliation{Department of Physics, King Mongkut's University of Technology Thonburi (KMUTT), Bangkok, Thailand}
\affiliation{Theoretical and Computational Physics (TCP); and Theoretical and Computational Science Center (TaCS), KMUTT, Thailand}	

\author{Lunchakorn Tannukij 
\footnote{Email: l\_tannukij@hotmail.com}}
\affiliation{Department of Physics, School of Science, King Mongkut's Institute of Technology Ladkrabang, 1 Chalong Krung 1 Alley, Lat Krabang, Bangkok, 10520, Thailand}

\author{Pitayuth Wongjun
\footnote{Email: pitbaa@gmail.com}}
\affiliation{The Institute for Fundamental Study (IF), Naresuan University,\\ 99 Moo 9, Tah Poe, Mueang Phitsanulok, Phitsanulok, 65000, Thailand}

\begin{abstract}

The nonextensive nature of black holes is one of the most intriguing discoveries.
In fact, the black hole entropy is a nonextensive quantity that scales by its surface area at the event horizon.
In our work, we extend the thermodynamic phase space of black holes by treating the nonextensive parameter analyzed via the R\'enyi entropy as the thermodynamic variable.
Using Euler's theorem for a homogeneous function of the black holes' mass, the compatible Smarr formula and the first law of black hole thermodynamics can be obtained.
It is also demonstrated that, by keeping the same form of the black hole mass, the R\'enyi temperature is straightforwardly defined as proposed in the literature. 
Since many different types of black holes can indeed be successfully treated with such a procedure, our consideration is fairly general.
It is worthwhile to argue that the black hole thermodynamics in R\'enyi statistics is rooted from the relation among geometric quantities in the same way as the standard approach corresponding to the Gibbs--Boltzmann statistics. 
Even though our results are based on classical gravity, they may pave the way to derive the R\'enyi temperature using the notion of quantum field in curved spacetime.

\end{abstract}
\maketitle{}
\newpage
\section{Introduction}\label{sect: intro}

Black holes and their physical properties have been one of the centers of interest to physics communities for decades. As mathematical solutions to general relativity, black holes are space-times whose curvatures are warped so that it is impossible for any object including light to escape once they enter the event horizons. 
Remarkably, the Event Horizon Telescope (EHT) collaboration showed the first image of the black hole's shadow in the center of galaxy M87 
\cite{EventHorizonTelescope:2019dse, EventHorizonTelescope:2019uob, EventHorizonTelescope:2019jan, EventHorizonTelescope:2019ths, EventHorizonTelescope:2019pgp, EventHorizonTelescope:2019ggy}. This pioneering success serves as solid evidence that black holes do exist in our universe and physically affect it so that the universe is just like what it is nowadays. 
One of the roles they must play is that, as thermal systems, black holes must be emitting thermal radiation similarly to any other ordinary matter in the universe. On the other hand, since black holes are known to be so strongly gravitating from which even light cannot escape, there should not be thermal radiation from black holes. Remarkably, Hawking demonstrated through quantum field theory in a curved space that the so-called Hawking radiation can be emitted from quantum phenomena around the even horizons \cite{Hawking:1975vcx}. His work paved a way for numerous studies on black hole thermodynamics and allowed physicists to understand such strongly-gravitating objects as thermal bodies.

Hawking's classical area theorem 
 states that the area of the event horizon always increases in the framework of classical general relativity \cite{Hawking:1971tu}. This notion 
led Bekenstein to postulate that the area of the event horizon and the surface gravity could be regarded as the entropy and temperature of a black hole, respectively \cite{Bekenstein:1973ur}. Later on, Bardeen et al. derived the laws of black hole mechanics that govern small perturbations in the black hole geometry \cite{Bardeen:1973gs}. 
By taking Bekenstein's postulate and the laws of black hole mechanics into account, they seem to imply a 
mathematical equivalence between the black hole mechanic laws and the thermodynamic laws. However, the entropy of the black hole can be identified as a nonextensive quantity since it is proportional to the area of the corresponding black hole. Therefore, the thermodynamics of a black hole may lie outside the validity of conventional extensive and additive Gibbs--Boltzmann (GB) entropy. As a result, the functional form of the black hole's entropy should be modified to correspond with the nonextensive nature of the black hole.

R\'enyi entropy is one of the entropies describing the equilibrium thermodynamics of a nonextensive system \cite{Renyi1959}. 
Moreover, it satisfies the zeroth law of thermodynamics so that the empirical temperature can be defined suitably \cite{Cimdiker:2022ics}. 
These properties of R\'enyi entropy motivate a number of investigations on the thermodynamics of black holes as nonextensive thermal systems 
\cite{Biro:2013cra,Czinner:2015eyk,Czinner:2017tjq,Promsiri:2021hhv,Alonso-Serrano:2020hpb,Promsiri:2020jga,Tannukij:2020njz,Nakarachinda:2021jxd,Sriling:2021lpr,Promsiri:2022qin,Chunaksorn:2022whl}. 
One of the worthwhile results is that nonextensivity manifests the thermodynamic stability of the black hole. 
A comprehensive discussion of how the Renyi entropy treatment can lead to stable configurations of black holes can be found in the literature \cite{Biro:2013cra,Czinner:2015eyk,Czinner:2017tjq,Promsiri:2021hhv,Alonso-Serrano:2020hpb,Promsiri:2020jga,Tannukij:2020njz,Nakarachinda:2021jxd,Sriling:2021lpr,Promsiri:2022qin,Chunaksorn:2022whl}.
Remarkably, the thermodynamics of black holes in an asymptotically flat and de Sitter (dS) spacetimes in the R\'enyi statistics can be in thermal equilibrium with the heat bath without embedding the black holes in the anti-de Sitter (AdS) spacetime or a finite fancy box.
However, in order to realize the corresponding thermodynamics based on the R\'enyi entropy, R\'enyi temperature is defined by assuming that the first law compatible with R\'enyi entropy retains the same form as that of standard GB entropy
\cite{Biro:2013cra,Czinner:2015eyk,Czinner:2017tjq,Promsiri:2021hhv,Alonso-Serrano:2020hpb,Promsiri:2020jga,Tannukij:2020njz,Nakarachinda:2021jxd,Sriling:2021lpr,Promsiri:2022qin,Chunaksorn:2022whl}. Since the first law does involve other thermodynamic quantities apart from the R\'enyi entropy and its conjugate temperature, one may need to assume a definition of a thermodynamic volume and its conjugate pressure (see Refs.~\cite{Promsiri:2020jga, Promsiri:2021hhv} for example), including those for other possible thermodynamic variables.
Such assumptions inevitably introduce inconsistencies among thermodynamic quantities since the assumed quantities do not give rise to a well-defined Smarr formula as discussed in Ref.~\cite{Sriling:2021lpr}. Conversely, the black hole thermodynamics based on GB statistics is well-equipped with its first law of thermodynamics which is able to be derived from the gravitational description from general relativity, including its well-defined Smarr formula which motivates consistent definitions for each thermodynamic quantity.
In fact, it is pointed out that, in order to relate the gravitational description to the black hole thermodynamics, the black hole temperature needs to be in the form of the Hawking temperature while the mass is kept to be the Arnowitt--Deser--Misner (ADM) mass \cite{Nojiri:2021czz}. For the R\'enyi entropy, the temperature is modified according to the nonextensivity, the form of the black hole mass is needed to be modified and then the modified mass must satisfy other gravitational theories rather than general relativity (GR) \cite{Nojiri:2022sfd}.  Actually, this inconsistency is rooted from the fact that the first law of black hole thermodynamics is not related to the gravitational description.

In this work, we demonstrate that 
it is possible to derive the first laws of thermodynamics compatible with the R\'enyi entropy from the Smarr formulae which are then derived from the gravitational description. Having the Smarr formulae, one is then able to define other thermodynamic quantities such as volume and pressure in a consistent manner. We also show that it is possible to apply the aforementioned treatments across different classes of black holes such as the Schwarzschild (Sch) black hole, Reissner--Nordstr\"om (RN) black hole, Kerr black hole, Schwarzschild--anti de Sitter/de Sitter (Sch-AdS/dS) black hole, and even the so-called de Rham--Gabadadze--Tolley (dRGT) black hole from the renowned dRGT massive gravity model.
Note that the thermodynamic phase space of the black holes in the R\'enyi description is constructed without modifying the black holes' mass. 
Therefore, this will provide connections between the black hole mechanical laws and the thermodynamic laws based on R\'enyi statistics, and allow concrete analyses on R\'enyi thermodynamics of the black hole.  
Despite the fact that our findings are based on classical gravity, this study may pave the way for deriving the R\'enyi temperature via the concept of quantum field in curved spacetime.

This work is organized as follows. 
Sec.~\ref{sec:GB} is devoted to reviewing how the Smarr formula is constructed from the relation of the geometric quantities based on Komar's integral. 
The approach of Euler’s theorem for a homogeneous function is also introduced for obtaining the equation for the first law.
The extended thermodynamic phase space for a black hole compatible with R\'enyi entropy is constructed by using the same procedure in Sec.~\ref{sec: Renyi}.
Various types of black holes are successfully treated, e.g., Sch and Kerr black holes.
Appendix~\ref{appen} shows the important expressions of the thermodynamic quantities for the RN, Sch-AdS/dS, and dRGT black holes.
In Sec.~\ref{sec: concl}, we have summarized and discussed the key points of this work.

\section{Thermodynamics with GB entropy}\label{sec:GB}

In this section, we will derive the Smarr formula and the first law of black hole mechanics using the GB statistics as it is instructive to investigate one for the R\'enyi case. 
In order to derive such a formula, it is worthwhile to review the suitable definition of the black hole mass and angular momentum. 
One of the useful definitions of the black hole mass and angular momentum is based on Komar's notion of the so-called Komar mass and angular momentum. 
For stationary and axially symmetric spacetimes, the Komar mass and angular momentum can be respectively given by \cite{Komar:1958wp, JKatz1985} (also see \cite{Poisson:2009pwt, Padmanabhan:2010zzb})
\begin{eqnarray}
	M &=& -\frac{1}{8\pi G}\oint_{\infty}\nabla^\mu K^\nu_{(t)} \text{d} S_{\mu\nu},\label{Komar M gen}\\
	J &=& \frac{1}{16\pi G}\oint_{\infty}\nabla^\mu K^\nu_{(\phi)} \text{d} S_{\mu\nu},\label{Komar J gen}
\end{eqnarray}
where $K^\nu_{(t)}$ and $K^\nu_{(\phi)}$ are the timelike and rotational Killing vectors of the spacetimes, respectively, and hence both of them satisfy the Killing's equations, $\nabla_{(\mu} K_{\nu)}=0$ where the bracket of indices is the symmetric permutation notation.  
For the two-dimensional spacelike hypersurface $S_{\mu\nu}$ described by the metric $\sigma_{AB}$ with coordinates $\theta^A$ (the indices $A$ and $B$ run over the dimensions of this two-surface), its surface element can be expressed as $\text{d}S_{\mu\nu}=2 n_{[\mu} r_{\nu]}\sqrt{\sigma} \text{d}^2\theta$ where $\sigma$, $n_\mu$ and $r_\mu$ are the determinant of $\sigma_{AB}$, timelike and spacelike vectors which are normal to this surface, respectively. 
Note that the integration of Komar expressions in Eqs.~\eqref{Komar M gen}~and~\eqref{Komar J gen} are the integral over the closed two-surface at infinity. 
This integral can also be split into two parts which are the inner integral (denoted by the subscript $H$) over the closed surface at the black hole horizon and the outer integral (denoted by the subscript $T$) for the shell between the black hole horizon to infinity. 
Applying Stokes' theorem to the later part, the mass and angular momentum are written as 
\begin{eqnarray}
       M &=& M_H +M_T 
       =  -\frac{1}{8\pi G}\oint_{r_H} \nabla^\mu K^\nu_{(t)} \text{d} S_{\mu\nu}-2 \int_{r_H}^\infty \text{d}\Sigma^\mu K^\nu_{(t)} \left(T_{\mu\nu} -\frac{1}{2} T g_{\mu\nu}\right),\\
        J &=& J_H + J_T 
        = \frac{1}{16\pi G}\oint_{r_H} \nabla^\mu K^\nu_{(\phi)} \text{d} S_{\mu\nu} - \int_{r_H}^\infty \text{d}\Sigma^\mu K^\nu_{(\phi)} \left(T_{\mu\nu} -\frac{1}{2} T g_{\mu\nu}\right),   
\end{eqnarray}
where $r_H$ and $\Sigma^\mu$ are the radius of the black hole's event horizon and the three-hypersurface (bounded by $S_{\mu\nu}$), respectively.
 and $T_{\mu\nu}$ is the energy-momentum tensor of the matter outside the black hole with the trace $T=T^\mu_{\,\,\,\,\mu}$.
Obviously, for the above total mass $M$ and angular momentum $J$, there are two contributions; the black hole and matter (outside the black hole). 
It is noted that these definitions of the black hole mass and angular momentum provide identical values to the ADM definition \cite{Arnowitt:1962hi}. 
It is very important to note that both Komar and ADM masses are identical only for the asymptotically flat black holes. 
It will be seen below that, for instance, the mass for Sch-AdS/dS black hole defined via Eq.~\eqref{Komar M gen} is not necessarily equal to the ADM mass.
In fact, there exists an infinite part in the Komar mass obtained from the matter contribution outside. 
One might treat this infinity as an effective mass or energy affected by observing the mass at asymptotically non-flat spacetime. 
As a result, by subtracting the effective part, the Komar mass is eventually finite and also the same as ADM one.

Interestingly, the geometric quantities are related via the well-known Smarr formula \cite{Smarr:1972kt}. 
It will be seen that this formula can be obtained by using the Komar expressions. 
For concreteness, let us consider the Kerr black hole. Since it is the solution in vacuum ($M_T = J_T = 0$), one then has $M=M_H$ and $J=J_H$. The linear combination of the black hole mass and angular momentum can be written as 
\begin{eqnarray}
       M - 2 \Omega_H J
       = -\frac{1}{8\pi G}\oint_{r_H}\nabla^\mu\left[K^\nu_{(t)}+\Omega_HK^\nu_{(\phi)}\right]\text{d}S_{\mu\nu}
       = \frac{\kappa}{4\pi G}\oint_{r_H}\sqrt{\sigma}\text{d}^2\theta 
       = \frac{\kappa}{4\pi G} A,  
\end{eqnarray}
where $\Omega_H$, $\kappa$, and $A$ are the angular velocity, surface gravity, and the surface area at the (outer) event horizon of the black hole, respectively. 
Here, we have used the fact that, for any stationary black hole in asymptotically flat spacetime, the event horizon is a Killing horizon \cite{Hawking:1973uf}.
Note that the Killing vector at the horizon is $K^\mu = K^\mu_{(t)} + \Omega_H K^\mu_{(\phi)}$. 
In other words, the normal vector of the null hypersurface is the mentioned Killing vector. 
By identifying that the black hole temperature and entropy are proportional to the surface gravity and surface area at the event horizon, respectively, one obtains the Smarr formula for the Kerr black hole as   
\begin{eqnarray}
	M = 2 T_\text{H} S_\text{BH} + 2 \Omega_H J,  \label{Smarr-Kerr}
\end{eqnarray}
where $T_\text{H}=\kappa/2\pi$ and $S_\text{BH}=A/(4G)$ are well-known as the Hawking temperature and Bekenstein--Hawking (BH) entropy, respectively. 
The aforementioned proportional constants are discovered from the thermal radiation of the black hole \cite{Bekenstein:1973ur, Hawking:1975vcx}.
It is noted that a black hole is a physically thermal system that has temperature and entropy; it is not just a system that is simply described by a convenient analogy with thermodynamics.
Moreover, for the static case, there is no angular momentum ($J=0$) so that the Smarr formula reduces to
\begin{eqnarray}
	M = 2 T_\text{H} S_\text{BH},\label{Smarr-Sch}
\end{eqnarray}	
which is the formula for the Sch black hole. 
In fact, in this case, the Killing vector at the horizon is timelike $K^\mu = K^\mu_{(t)}$. 
It is very important to note that the mass in Eq.~\eqref{Smarr-Sch} is the mass contributed from the black hole part $M_H$ for any black holes in asymptotically flat spacetime.

Let us further discuss the case of non-vacuum case, i.e., $M_T \neq 0$. 
For this case, it is instructive to consider the non-rotating ($J=0$) charged black hole (RN black hole) in which the metric can be written as 
\begin{eqnarray}
	\text{d}s^2 &=& -f(r) \text{d}t^2 + f^{-1}(r) \text{d}r^2 + r^2 \text{d}\Omega^2,\label{ds^2}\\
	f(r) &=& 1- \frac{2GM_\text{ADM}}{r} + \frac{q^2}{r^2}, \label{ds-RN}
\end{eqnarray}
where $M_\text{ADM}$ is the ADM mass of the black hole. 
For the metric of any static and spherically symmetric black holes taken in the form of Eq.~\eqref{ds^2}, this mass can be obtained in terms of $r_H$ from solving the horizon equation, $f(r_H)=0$. 
$q$ is the electric charge of the black hole and $\text{d}\Omega^2 = \text{d}\theta^2 +\sin^2\theta \text{d}\phi^2$. 
The energy-momentum tensor contributed from the electric charge can be written as 
\begin{eqnarray}
	T^\mu_{\,\,\,\,\,\nu} = F^{\mu\rho} F_{\nu\rho} - \frac{1}{4} F_{\rho\sigma}F^{\rho\sigma} \delta^\mu_\nu,
\end{eqnarray}
where $F_{\mu\nu} = \nabla_\mu A_\nu -\nabla_\nu A_\mu$ and $A_\mu$ is the $U(1)$ gauge field. 
The mass due to the matter contribution outside the black hole can be expressed as
\begin{eqnarray}
	M_T
	&=&-2 \int_{r_H}^\infty \text{d}\Sigma_\mu K^\nu_{(t)} \left(T^\mu_{\,\,\,\,\,\nu} -\frac{1}{2} T \delta^\mu_\nu\right)\nonumber\\
    &=&2\int_{r_H}^\infty n_\mu\sqrt{h}\text{d}^3yK^\nu_{(t)} \left(T^\mu_{\,\,\,\,\,\nu} -\frac{1}{2} T \delta^\mu_\nu\right)\nonumber\\
	&=&-q \int_{r_H}^\infty \text{d}r \frac{q}{4\pi r^2}\nonumber\\
	&=& q \Phi_q.
\end{eqnarray}
where $h$ is the determinant of the metric $h_{ij}$ with coordinates $y^i$ describing the three-surface $\Sigma^\mu$, and $\Phi_q$ is the electric potential of the black hole, $\Phi_q=q/(4\pi r_H)$.
Note that $n_\mu$ is the timelike normal vector for the spacelike hypersurface $\Sigma$.
Here, we have used the Killing vector $K^\mu_{(t)}$ and normal vector $n_\mu$ as $K^\mu_{(t)}=\delta^\mu_0$ and $n_\mu=-f^{-1/2}\delta^0_\mu$. 
The (00) component and the trace of this energy-momentum tensor are straightforwardly computed as $T^0_{\,\,\,\,0} = -\frac{q^2}{32\pi^2r^4}$ and $T=0$, respectively.
As a result, the Smarr formula for the RN black hole reads  
\begin{eqnarray}
	M = M_\text{ADM}= M_H + M_T = 2T_\text{H}S_\text{BH} + q \Phi_q. \label{Smarr-RN}
\end{eqnarray}
Note that we have used the fact that the surface gravity for static and spherically symmetric solutions taken in the form of Eq.~\eqref{ds^2} can be computed as $\kappa=\frac{1}{2}\partial f/\partial r|_{r=r_H}$.
Furthermore, by including the rotating effect, e.g., axially symmetric solutions, it can also be generalized to the case of the charged rotating black hole or Kerr--Newman (KN) black hole as 
\begin{eqnarray}
	M = 2T_\text{H}S_\text{BH} +2\Omega_H J+ q \Phi_q.
\end{eqnarray}
Let us emphasize here that this formulation is useful and can be applied in other kinds of black holes with asymptotically flat spacetime \cite{Poisson:2009pwt, Padmanabhan:2010zzb}.

Now, moving our attention to the case of black holes in non-asymptotically flat spacetime, e.g. in asymptotically AdS/dS spacetime, it is found that the mass and angular momentum evaluated at spatial infinity are divergent. In order to obtain a suitable form of the black hole mass, there have existed many methods found in the literature \cite{Magnon:1985sc,Kastor:2008xb,Kastor:2009wy,Caldarelli:1999xj,Awad:1999xx,Das:2000cu,Hajian:2015xlp,Hajian:2016kxx,Astorino:2016hls,Gao:2023luj,Peng:2020cfy}. For example, the divergent term can be eliminated by subtracting the suitable choice of the reference background \cite{Magnon:1985sc}. Moreover, the black hole mass can be generalized by introducing an additional term related to an anti-symmetric Killing potential to cancel the divergent term \cite{Kastor:2008xb,Kastor:2009wy}. In this work, we will use the similar idea proposed in Ref.~\cite{Magnon:1985sc}. 
By subtracting a contribution related to the reference background, the divergent term is cancelled.
Moreover, it is found that such the contribution provides the Killing potential term as found in Refs.~\cite{Kastor:2008xb,Kastor:2009wy}.
To see explicitly, let us consider the Sch-AdS/dS black hole with the horizon function given by
\begin{eqnarray}
	f(r) &=& 1- \frac{2GM_\text{ADM}}{r} - \frac{\Lambda r^2}{3},
\end{eqnarray}
where $\Lambda$ is the cosmological constant which is negative (positive) for the Sch-AdS(dS) black hole.
Note that the effective energy-momentum tensor in this case is $T^\mu_{\,\,\,\,\nu}=-\frac{\Lambda}{8\pi G}\delta^\mu_\nu$. 
As a result, the two contributions in Komar mass can be evaluated as
\begin{eqnarray}
	M_H 
	&=& \frac{1}{8\pi G}\oint_{r_H} \left(\frac{2GM_\text{ADM}}{r^2}-\frac{2\Lambda r}{3 }\right) r^2 \sin \theta \text{d}\theta \text{d}\phi
	= M_\text{ADM} - \frac{\Lambda }{3 G } r^3_H,\\
	M_T 
	&=& -2\int_{r_H}^\infty\left(\frac{\Lambda}{8\pi G}\right)r^2\sin\theta\text{d}r\text{d}\theta\text{d}\phi
	= - \frac{\Lambda}{3G}r^3\Big|_{r\to\infty}+\frac{\Lambda}{4\pi G} \frac{4\pi}{3} r^3_H.
\end{eqnarray}
The Komar mass of the black hole reads
\begin{eqnarray}
	M=M_H+M_T
    =M_\text{ADM}- \frac{\Lambda}{3G}r^3\Big|_{r\to\infty}.
\end{eqnarray} 
Obviously, the additional (infinite) term to the ADM mass comes from the contribution of matter outside the black hole. Let us define this term by
\begin{eqnarray}
	M_\infty=-\frac{\Lambda}{3G}r^3\Big|_{r\to\infty}.
\end{eqnarray}
As mentioned previously, this term might be interpreted as the mass or energy affected by measuring the mass of the black hole by an observer who stays at non-asymptotically flat spacetime. 
In addition, the expression of $M_\infty$ can be obtained from computing the Komar integral by Eq.~\eqref{Komar M gen} of the pure AdS/dS space. 
Therefore, this infinite term is also interpreted as background energy due to the existence of asymptotically AdS/dS space.
Then, the proper mass is redefined by subtracting such term from the original Komar mass as follows
\begin{eqnarray}
	M\to M-M_\infty
    =M_\text{ADM}
	=\frac{r_H}{2G}\left(1-\frac{\Lambda}{3}r_H^2\right)
	=2\left(\frac{1-\Lambda r_H^2}{4\pi r_h}\right)\frac{\pi r_H^2}{G}
	-2\left(-\frac{\Lambda}{8\pi G}\right)\frac{4\pi}{3},\label{M sub}
\end{eqnarray}
where $M_\text{ADM}$ is obtained by solving the horizon condition $f(r_H)=0$.

As a result, the Smarr formula for the Sch-AdS/dS black hole can be written as 
\begin{eqnarray}
	M = 2 T_\text{H} S_\text{BH} - 2\left(-\frac{\Lambda}{8\pi G}\right) \frac{4\pi}{3} r^3_H.\label{Smarr Sch-AdS/dS}
\end{eqnarray}
One further notices that the second term on the right-hand side of Eq.~\eqref{Smarr Sch-AdS/dS} is proportional to $r_H^3$ or in the dimension of three-volume. 
It is useful to interpret the contribution from $\Lambda$ as a thermodynamic pressure of the system since it manifests as a conjugate variable to the volume.
By defining the pressure as $P =-\Lambda/(8\pi G)$, the Smarr formula for the AdS/dS black hole becomes 
\begin{eqnarray}
	M = 2 T_\text{H} S_\text{BH} - 2 P V. \label{Smarr Sch-AdS/dS final}
\end{eqnarray}
Note that we can use the same strategy in order to obtain further generalization. 
For example, for the charged black hole in asymptotically AdS/dS space, the Smarr formula can be written as $M = 2 T_\text{H} S_\text{BH} -2 P V + q\Phi_q$.

It is important to note that the Smarr formula for non-asymptotically flat spacetime can be obtained by employing the Killing potential $\omega^{\nu\mu}$ \cite{Kastor:2008xb,Kastor:2009wy}. By using the relation to defining the Killing potential $K^{\mu} _{(t)}= \nabla_\nu \omega^{\nu\mu}$, our subtracting term can be written in terms of the Killing potential as 
\begin{eqnarray}
	M_\infty &=& \frac{-1}{4 \pi G} \int_\infty \text{d}\Sigma_\mu K^{\nu} _{(t)} R^\mu_{\,\,\,\nu}\nonumber\\ \nonumber
                 &=& \frac{\Lambda}{4 \pi G} \int_\infty \text{d}\Sigma_\mu K^{\nu} _{(t)} \delta^\mu_\nu\\ \nonumber
                &=& \frac{\Lambda}{4 \pi G} \int_\infty \text{d}\Sigma_\mu \nabla_\nu \omega^{\nu\mu}\\ 
                &=& \frac{-\Lambda}{8 \pi G} \oint_\infty \text{d}S_{\mu\nu} \omega^{\mu\nu}.
\end{eqnarray}
Here, we have used $R^\mu_{\,\,\,\nu}=-\Lambda\delta^\mu_\nu$ in the second line and applied Stoke's theorem in the last line.
Note that the contribution of this term at $r_H$ provides $M_T$ in our notation. One can see that the black hole mass in our approach defined in Eq.~(\ref{M sub}) is actually the same as one found in the literature. 

Moreover, it is found that one can use this strategy not only for black holes in GR but also for those in modified gravity theories. 
For example, for the black hole in the dRGT massive gravity theory (called dRGT black hole), the horizon function can be written as \cite{Ghosh:2015cva, Boonserm:2017qcq}
\begin{eqnarray}
	f(r) &=& 1 - \frac{2GM_\text{ADM}}{r} - m_{g}^{2} (c_{2}r^{2} - c_{1}r - c_{0}),
\end{eqnarray}
where $m_g, c_2, c_1$ and $c_0$ are model parameters. 
The components of the effective energy tensor are given by $T^0_{\,\,\,\,0}=T^1_{\,\,\,\,1}=\frac{m_g^2}{8\pi G}\left(c_2+\frac{c_1}{r}+\frac{c_0}{r^2}\right)$ and $T^2_{\,\,\,\,2}=T^3_{\,\,\,\,3}=\frac{m_g^2}{8\pi G}\left(3c_2+\frac{c_1}{r}\right)$. 
The proper black hole mass can be defined in the same way as done in the Sch-AdS/dS case. The infinite term for the dRGT black hole is given by
\begin{eqnarray}
	M_\infty=\left[2\int\frac{m_g^2}{8\pi G}\left(3c_2+\frac{c_1}{r}\right)r^2\sin\theta\text{d}r\text{d}\theta\text{d}\phi\right]\bigg|_{r\to\infty}
	=\frac{m_g^2}{3G}\left(3c_2r^3+\frac{3c_1}{2}r^2\right)\bigg|_{r\to\infty}.
\end{eqnarray}
Eventually, the finite mass for this black hole can be written in the form of the Smarr formula as 
\begin{equation}
	M-M_\infty = M_\text{ADM} = 2 T_\text{H} S_\text{BH} - 2PV  + 2c_{0} \Phi_{0} + c_{1} \Phi_{1},\label{Smarr dRGT}
\end{equation}
where the pressure is identified as $P \equiv \frac{3}{8 \pi G} m_{g}^{2}$. 
The conjugate volume is obtained as $V=\frac{4\pi}{3}(c_0r_H+c_1r_H^2-c_2r_H^3)$. 
Moreover, we also treated the parameters $c_0$ and $c_1$ as thermodynamic variables. 
Their conjugated potentials are given by as $\Phi_0 = \frac{4 \pi}{3} P r_H$ and $\Phi_1 = \frac{4 \pi}{3} P r_H^2$.
It is noticed that the mentioned pressure has the opposite sign to that for the Sch-dS case.
This means that, in asymptotically dS spacetime, $P$ and $V$ for the dRGT black hole can be concurrently positive while those for Sch-dS black hole cannot \cite{Chunaksorn:2022whl}. 
Note that there exist other ways to identify thermodynamic variables for the dRGT black hole (see \cite{Xu:2015rfa, Hendi:2017fxp} for examples).

From the above discussion, one can see that the Smarr formulae for the black holes are rooted from the description of classical gravitation. 
In other words, these relations are purely obtained from the geometrical quantities of the spacetimes.
Identifying the proper thermodynamic quantities, it serves as a thermodynamic formulation for the black hole. 
Therefore, the thermodynamic equations are supposed to be derived from the Smarr formula via the Legendre structure.
In the following discussions, we will show how the first law of thermodynamics is obtained from the Smarr formula using Euler's theorem for a homogeneous function.

A function $f(x^I)$ is said to be a homogeneous function of degree $k$ in $x^I$ if the function satisfies $f(a x^I) = a^k f(x^I)$ for every nonzero scalar $a$. 
Considering Euler's theorem for a homogeneous function, it states that if $f(x^I)$ is a homogeneous function of degree $k$, this function satisfies the partial differential equation as follows
\begin{equation}
	k f(x^I) = x^I \frac{\partial f}{\partial x^I}.\label{Smarr for homo fn}
\end{equation} 
The differentiation of this homogeneous function simply reads
\begin{eqnarray}
	\text{d}f = \frac{\partial f}{\partial x^I}\text{d}x^I.\label{1st law for homo fn}
\end{eqnarray}
Note that the Einstein summation convention has been used in Eqs.~\eqref{Smarr for homo fn}~and~\eqref{1st law for homo fn}.
Now, let us apply this theorem to the Kerr black hole for an instructive consideration. 
For the Kerr black hole, its mass can be expressed in terms of the BH entropy and angular momentum as follows
\begin{equation}
	M
	=\sqrt{\frac{4\pi^2J^2+S_\text{BH}^2}{4\pi GS_\text{BH}}}.
\end{equation}
It is noticed that the black hole mass $M(S_\text{BH}, J)$ is indeed the homogeneous function degree $1/2$, i.e., $M(a S_\text{BH}, a J)=a^{1/2} M(S_\text{BH}, J)$. 
According to Euler's theorem for the homogeneous function, the mass and its derivatives obey the differential equation~\eqref{Smarr for homo fn} which becomes
\begin{equation}
	\frac{1}{2} M = S_\text{BH} \left(\frac{\partial M}{\partial S_\text{BH}}\right)_J + J \left(\frac{\partial M}{\partial J}\right)_{S_\text{BH}}.\label{diff eq Kerr}
\end{equation}
The conjugate thermodynamic variables of the entropy and angular momentum can be defined as 
\begin{equation}
	T = \left(\frac{\partial M}{\partial S_\text{BH}}\right)_J, \quad 
	\Omega=\left(\frac{\partial M}{\partial J}\right)_{S_\text{BH}}.
\end{equation}
It has been found that these conjugate quantities $T$ and $\Omega$ are identical to the Hawking temperature ($T=T_\text{H}$) and angular velocity ($\Omega=\Omega_H$) of the black hole, respectively. 
Therefore, the differential equation~\eqref{diff eq Kerr} is indeed the Smarr formula~\eqref{Smarr-Kerr}.
As a result, the infinitesimal differentiation of the black hole mass reads 
\begin{equation}
	\text{d} M = \left(\frac{\partial M}{\partial S_\text{BH}}\right)_J\text{d}S_\text{BH} + \left(\frac{\partial M}{\partial J}\right)_{S_\text{BH}}\text{d}J
	 = T_\text{H} \text{d}S_\text{BH} + \Omega_H \text{d}J.
\end{equation}
Emphasize that it is actually equivalent to the first law of thermodynamics.
In other words, this black hole system is supposed to be described by equilibrium thermodynamics.
The Hawking temperature $T_\text{H} = \kappa/(2\pi)$ characterizes a system that has minimum energy and maximum entropy. 
Therefore, the temperature defined in this way is compatible with the zeroth law of thermodynamics.

For non-vacuum black hole solutions, we can also use this strategy in order to obtain the first law. For this case, it is instructively to consider the RN black hole solution where the metric is expressed in Eq.~\eqref{ds-RN}.  From this solution, the black hole mass, charge, and entropy obey the relation 
\begin{equation}
	M 
	=\frac{1}{2\sqrt{\pi G S_\text{GB}}}\left(S_\text{BH}+\frac{\pi}{G}q^2\right).
\end{equation}
Therefore, this mass $M(S_\text{GB},q^2)$ is the homogeneous function degree $1/2$, i.e., $M(a S_\text{GB}, a q^2) = a^{1/2} M(S_\text{GB}, q^2)$. By using Euler's theorem for the mass function as considered in the case of the Kerr black hole, we obtain
\begin{eqnarray}
	\frac{1}{2} M &=& S_\text{BH} \left(\frac{\partial M}{\partial S_\text{BH}}\right)_q + q^2\left(\frac{\partial M}{\partial q^2}\right)_{S_{BH}},
\end{eqnarray}
which leads to the Smarr formula~\eqref{Smarr-RN} by defining the Hawking temperature and electric potential for the RN black hole as follows
\begin{eqnarray}
	T_\text{H}=\left(\frac{\partial M}{\partial S_\text{BH}}\right)_q,\quad
	\Phi_q=\left(\frac{\partial M}{\partial q}\right)_{S_{BH}}.
\end{eqnarray}
In the same way, as done in the Kerr black hole case, the equation associated with the first law of thermodynamics for the RN black hole can be written as
\begin{equation}
	\text{d} M = \left(\frac{\partial M}{\partial S_\text{BH}}\right)_q \text{d}S_\text{BH} +  \left(\frac{\partial M}{\partial q}\right)_{S_{BH}} \text{d}q 
	= T_\text{H}\text{d}S_\text{BH} + \Phi_q \text{d}q.
\end{equation}

From these investigations, one can see that it is possible to follow the same procedure to obtain the first law for other asymptotically flat black holes, e.g., the KN black hole.
It is very important to note that this procedure is also applicable to black holes in non-asymptotically flat spacetime.
For instance, let us consider the Sch-AdS/dS black hole with the mass taken in the terms of the entropy and pressure as 
\begin{eqnarray}
	M=\sqrt{\frac{S_\text{BH}}{\pi G}}\left(\frac{1}{2}+\frac{4}{3}G^2S_\text{BH}P\right).
\end{eqnarray}
Hence, the black hole mass $M(S_\text{BH},P^{-1})$ is the homogeneous function degree $1/2$, i.e., $M(a S_\text{BH}, a P^{-1}) = a^{1/2} M(S_\text{BH}, P^{-1})$. The differential equation corresponding to the Smarr formula~\eqref{Smarr Sch-AdS/dS final} is obtained as
\begin{eqnarray}
	\frac{1}{2}M = S_\text{BH} \left(\frac{\partial M}{\partial S_\text{BH}}\right)_P + P^{-1}\left(\frac{\partial M}{\partial P^{-1}}\right)_{S_{BH}},
\end{eqnarray}
where 
\begin{equation}
	T_\text{H} = \left(\frac{\partial M}{\partial S_\text{BH}}\right)_P, \quad 
	V = \left(\frac{\partial M}{\partial P}\right)_{S_\text{BH}}.
\end{equation}
Then, the first law can be written as
\begin{equation}
	\text{d} M = T_\text{H} \text{d}S_\text{BH} + V \text{d}P.
\end{equation}
According to the form of this differentiation, the mass of the Sch-AdS/dS black hole is interpreted as the enthalpy rather than the internal energy of the system. 
Note that for the Sch-dS case, the thermodynamic variable $P =-\Lambda/(8\pi G)$ is negative and may be interpreted as the tension rather than the pressure of the system. 
Moreover, it is possible to have two horizons experienced by an observer; the black hole horizon $r_b$ and cosmic horizon $r_c$. 
The system for each horizon can be individually defined so that the first laws are expressed as 	\begin{eqnarray}	
	\text{d} M = \pm T_{b,c} \text{d}S_{b,c} + V_{b,c} \text{d}P, \label{1st law-AdS/dS}
\end{eqnarray}
where the plus (minus) sign and the subscript $b$ $(c)$ stand for the thermodynamic system for the black hole (cosmic) horizon.
It is seen that these systems defined in this way still use the same thermodynamic variables $M$ and $P$.
In addition, the minus sign in front of the heat term in Eq.~\eqref{1st law-AdS/dS} is introduced from the fact that the temperature can be suitably defined as  $T_c = -\big(\partial M / \partial S_c\big)_P$ 
\cite{Dolan:2013ft, Kubiznak:2016qmn}
(see also \cite{Banihashemi:2022htw} for the discussion on the ambiguity of this minus sign). 
Straightforwardly, the thermodynamic relations for other generalized black holes, e.g., the RN-AdS/dS, KN, Kerr-AdS/dS, and KN-AdS/dS black holes, can be constructed.

As we have mentioned previously, the Smarr formula can be extended to apply to black holes in modified gravity theories. 
To realize the advantage of the procedure, Euler's theorem for the homogeneous function is still used for those black holes. 
We will consider the black hole in the dRGT massive gravity theory as an example. 
The black hole mass is given by
\begin{eqnarray}
	M=\sqrt{\frac{S_\text{BH}}{\pi G}}\left[\frac{1}{2}+\frac{4}{3}\left(\pi GPc_0+\pi^{1/2}G^{3/2}S_\text{BH}^{1/2}Pc_1-G^2S_\text{BH}Pc_2\right)\right].
\end{eqnarray}
From Euler's theorem for the homogeneous function degree $1/2$ 
\begin{eqnarray}
	M(aS_\text{BH},aP^{-1},ac_{0},ac_{1}^{2}) = a^{1/2} M(S_\text{BH},P^{-1},c_0,c_1^2),
\end{eqnarray}
the Smarr formula~\eqref{Smarr dRGT} corresponds to the differential equation
\begin{eqnarray}
	\frac{1}{2}M &=& 
	S_\text{BH} \left(\frac{\partial M}{\partial S_\text{BH}}\right)_{P,c_0,c_1} 
	+ P^{-1}\left(\frac{\partial M}{\partial P^{-1}}\right)_{S_\text{BH},c_0,c_1}\nonumber\\
	&&+ c_0\left(\frac{\partial M}{\partial c_0}\right)_{S_\text{BH},P,c_1}
	+ c_1\left(\frac{\partial M}{\partial c_1}\right)_{S_\text{BH},P,c_0},
\end{eqnarray}
where 
\begin{eqnarray}
	T_\text{H} = \left(\frac{\partial M}{\partial S_\text{BH}}\right)_{P,c_0,c_1}, &\quad& 
	V = \left(\frac{\partial M}{\partial P}\right)_{S_\text{BH},c_0,c_1},\nonumber\\
	\Phi_0 = \left(\frac{\partial M}{\partial c_0}\right)_{S_\text{BH},P,c_1},&\quad&
	\Phi_1 =  c_1\left(\frac{\partial M}{\partial c_1}\right)_{S_\text{BH},P,c_0}.
\end{eqnarray}
Eventually, the first law of the black hole in dRGT massive gravity theory can be written as
\begin{equation}
	\text{d}M = \pm T_{b,c}\text{d}S_{b,c} + V_{b,c}\text{d}P + \Phi_{0(b,c)}\text{d}c_{0} + \Phi_{1(b,c)}\text{d}c_{1},
\end{equation}
where the system for the cosmic horizon is taken into account.
Again, the plus (minus) sign and the subscript $b$ $(c)$ stand for the thermodynamic system for the black hole (cosmic) horizon.
Note also that $T_c=-(\partial M/\partial S_c)_{P, c_0, c_1}$.

From this investigation, one can see that the thermodynamic laws are somehow equivalent to the gravitational notion. 
Actually, there are the same equations if we can properly identify the thermodynamic variables from the gravitational quantities. 
Moreover, it is found that the procedure providing such equations is quite general since it can be applied to various kinds of black holes. 
At this stage, it is worthwhile to note that the thermodynamic equations are obtained by identifying the entropy as the surface area of the black hole's horizon, specifically $S_\text{BH} = A/(4G)$. 
Then, the temperature corresponding to the conjugate thermodynamic variable of the entropy coincides with the Hawking temperature. 
As a result, the mentioned thermodynamic systems are based on the GB statistics corresponding to the extensive system. 
However, it contradicts itself because the entropy of the black hole is not the extensive quantity which is proportional to area, not volume.

A possible way to clarify the mentioned issue is promoting the black hole entropy to be described by a nonextensive entropy, for example, the R\'enyi entropy \cite{Renyi1959}. 
Unfortunately, it is found that, in the literature, the first law associated with the R\'enyi entropy is adopted without a link to the gravitational description. 
In the next section, we will show that it is possible to obtain the equations corresponding to the first law from a gravitational description by using the same procedure as we reviewed in this section.


\section{Thermodynamics with R\'enyi entropy}\label{sec: Renyi}

One of the key ideas of the procedure reviewed in the previous section is that the black hole thermodynamic variables and their relations are obtained from the gravitational description. 
In other words, Euler's theorem for the homogeneous function corresponding to the geometric quantity, i.e., the black hole mass, is applied in order to obtain the Smarr formula and the equation for the first law of thermodynamics. 
It is very important to mention that in order to preserve the structure of gravitational description, the Smarr formula should be maintained in the same form.

In this section, we are interested in modifying the role of the homogeneous function to satisfy the nonextensive nature of entropy. 
It will be seen that the black holes' Smarr formulae and equations for the first laws associated with nonextensivity analyzed via the R\'enyi entropy can be derived using this procedure.

To study the nonextensive nature of black holes, the BH entropy is supposed to be described by the Tsallis entropy $S_\text{T}$ \cite{Tsallis:1987eu}. 
According to the nonadditive composition rule of the Tsallis entropy, $S^{12}_\text{T}\neq S^1_\text{T}+S^2_\text{T}$, unfortunately, it leads to the zeroth-law incompatibility. 
A solution to deal with this problem is transforming the Tsallis entropy via the formal logarithm map \cite{Biro2011}. 
The resulting entropy is indeed the R\'enyi entropy \cite{Renyi1959}, $S_\text{R}=\frac{1}{\lambda}\ln(1+\lambda S_\text{T})$, where the parameter $\lambda$ plays roles of both formal logarithm map parameter and nonextensive parameter in the R\'enyi entropy. 
Hence, in this viewpoint, the black hole entropy is not the surface area at the horizon, but it is a logarithm function of the area instead as follows
\begin{eqnarray}
	S_\text{R} = \frac{1}{\lambda}\ln(1+\lambda S_\text{BH})
	= \frac{1}{\lambda} \ln\left(1+ \lambda \frac{A}{4G}\right).
\end{eqnarray}
Here, the R\'enyi entropy is indeed a one-parameter generalization of the GB entropy. It is important to note that there is not necessary to interpret the Bekenstein-Hawking entropy $S_{BH} = A/(4G)$ as the Tsallis entropy. It is sufficient to interpret the entropy in the above equation as a formal logarithm mapped entropy of the Bekenstein-Hawking entropy \cite{Biro2011}. Therefore, in this case, the above form of the entropy can be viewed as the parameter generalization of the Bekenstein–Hawking entropy. This interpretation does not alter the results in the following content.
In fact, the BH entropy is recovered by taking the limit $\lambda\to0$ to the R\'enyi one, $\displaystyle S_\text{BH}=\lim_{\lambda\to0}S_\text{R}$.
Straightforwardly, one also can write the BH entropy in terms of the R\'enyi one as follows
\begin{eqnarray}
	S_\text{BH}=\frac{1}{\lambda}\left(e^{\lambda S_\text{R}}-1\right).
\end{eqnarray}
By plugging the above expression into the black holes' mass shown in the previous section, the mass is now the function of the R\'enyi entropy.
Due to the existence of the additional nonextensive parameter in R\'enyi entropy, the thermodynamic phase space for the black hole should be extended. 
In other words, the nonextensive parameter must be treated as the thermodynamic variable for the black hole described by the R\'enyi entropy as will be seen soon.

Let us begin with considering a simple example such as the Sch black hole case. 
The black hole mass is written in terms of the R\'enyi entropy as
\begin{eqnarray}
	M =\frac{1}{2} \sqrt{\frac{e^{\lambda  S_R}-1}{\pi G \lambda }}.
\end{eqnarray}
From this expression, one obviously needs to interpret the nonextensive parameter as the thermodynamic variable in order to indicate that the mass is a homogeneous function of the R\'enyi entropy. 
As a result, the black hole mass is the homogeneous function degree $1/2$, 
\begin{eqnarray}
	M(a S_\text{R}, a \lambda^{-1} ) = a^{1/2} M( S_\text{R},  \lambda^{-1} ).
\end{eqnarray}
Applying the Euler's theorem for $M( S_\text{R},  \lambda^{-1} )$, we obtain the Smarr formula as
\begin{eqnarray}
	\frac{1}{2} M 
	= S_\text{R}\left(\frac{\partial M}{\partial S_\text{R}}\right)_\lambda + \lambda^{-1}\left(\frac{\partial M}{\partial \lambda^{-1}}\right)_{S_\text{R}} 
	= T_\text{R}S_\text{R} - \Phi_\lambda\lambda,
\end{eqnarray}
where the R\'enyi temperature can be defined as a conjugate variable to the entropy, $T_\text{R} = \big(\partial M /\partial S_\text{R}\big)_\lambda$ and the conjugate variable to the nonextensive parameter can be defined as $\Phi_\lambda = \big(\partial M /\partial \lambda\big)_{S_\text{R}}$. 
The explicit forms of both thermodynamic variables can be expressed as
\begin{eqnarray}
	T_\text{R} &=& \frac{1}{4\pi r_H} \left(1+ \frac{\lambda \pi r_H^2}{G}\right),\\
	\Phi_\lambda &=&\frac{1}{4\pi r_H \lambda^2}\left[\left(1+ \frac{\lambda \pi r_H^2}{G}\right) \ln \left(1+ \frac{\lambda \pi r_H^2}{G}\right) -\frac{\lambda \pi r_H^2}{G}\right].
\end{eqnarray}
The form of the R\'enyi temperature coincides with one found in literature \cite{Biro:2013cra, Czinner:2015eyk}. 
Moreover, in the limit $\lambda\to0$, the R\'enyi temperature reduces to the Hawking temperature while $\Phi_\lambda\lambda$ goes to zero. 
Therefore, the system in this limit recovers one described by GB statistics as expected. 
Then, the differentiation of $M( S_\text{R},  \lambda^{-1} )$ corresponding to the first law can be obtained as follows 
\begin{eqnarray}
	\text{d}M =T_\text{R} \text{d}S_\text{R} + \Phi_\lambda \text{d}\lambda.
\end{eqnarray}
Let us emphasize that the R\'enyi temperature can be defined to be consistent with R\'enyi entropy by keeping the black hole mass in the same form.
It is interestingly found that, for a small value of $\lambda$, the leading order of $\Phi_\lambda$ is proportional to $r_H^3$.
$\Phi_\lambda$ might be interpreted as the thermodynamic volume where the nonextensive parameter can play the role of pressure. 
By using dimensional analysis, the thermodynamic volume and pressure can be identified as 
\begin{eqnarray}
	V_\lambda &=&\frac{8 G^2 }{3\pi r_H \lambda^2}\left[\left(1+ \frac{\lambda \pi r_H^2}{G}\right) \ln \left(1+ \frac{\lambda \pi r_H^2}{G}\right) -\frac{\lambda \pi r_H^2}{G}\right],\\
	P_\lambda &=& \frac{3}{32 G^2} \lambda.
\end{eqnarray}
The Smarr formula and the first law for the re-interpretation respectively read
\begin{eqnarray}
	M  &=& 2T_\text{R} S_\text{R} - 2V_\lambda P_\lambda,\\
	\text{d}M  &=& T_\text{R} \text{d}S_\text{R} + V_\lambda \text{d}P_\lambda.
\end{eqnarray}
In this sense, the black hole mass can be interpreted as the thermodynamic enthalpy instead of internal energy. 
Moreover, one can see that the nonextensive parameter in Sch black hole can act in a similar role to the cosmological constant in Sch-AdS one.

By using this prescription, one can further construct the thermodynamics for the stationary and axial symmetric black holes, e.g., the Kerr black hole. 
As a result, the mass of the Kerr black hole can be expressed in terms of the R\'enyi entropy, nonextensive parameter, and angular momentum as 
\begin{eqnarray}
	M = \frac{1}{2\sqrt{\pi G}}  \sqrt{\frac{(e^{\lambda  S_\text{R}}-1)}{\lambda } +\frac{4\pi^2J^2 \lambda}{(e^{\lambda  S_\text{R}}-1)}}.
\end{eqnarray}
From this expression, the mass is the homogeneous function degree $1/2$ as 
\begin{eqnarray}
	M(a S_\text{R}, a \lambda^{-1},a J ) = a^{1/2} M( S_\text{R}, \lambda^{-1}, J ).
\end{eqnarray}
By following the same strategy as performed in the Sch case, the equations for the Smarr formula and the first law can be written as 
\begin{eqnarray}
	M &=& 2T_\text{R} S_\text{R} - 2\Phi_\lambda \lambda + 2\Omega_\text{R} J,\\
	\text{d}M &=& T_\text{R} \text{d}S_\text{R} + \Phi_\lambda \text{d}\lambda + \Omega_\text{R} \text{d} J,
\end{eqnarray}
where the conjugate variables are expressed as
\begin{eqnarray}
	T_\text{R} &=& \left(\frac{\partial M}{\partial S_\text{R}}\right)_{\lambda, J} 
	= \frac{\left(1-j^2\right) \left(\frac{A \lambda }{4 G}+1\right)}{2 \sqrt{\pi  A} \sqrt{j^2+1}},\\
	\Phi_\lambda &=& \left(\frac{\partial M}{\partial \lambda} \right)_{S_\text{R},J}
	= \frac{\left(1-j^2\right) \left[\left(\frac{A \lambda }{4 G}+1\right) \ln \left(\frac{A \lambda }{4 G}+1\right)-\frac{A \lambda }{4 G}\right]}{2 \sqrt{\pi  A} \lambda ^2 \sqrt{j^2+1}},\\
	\Omega_\text{R} &=& \left(\frac{\partial M}{\partial J}\right)_{S_\text{R},\lambda}
    =\frac{2j}{\sqrt{j^2+1}}\sqrt{\frac{\pi}{A}},
\end{eqnarray}
where $j^2=64 \pi ^2 G^2 J^2/A^2$.

For non-vacuum solutions, it is instructive to consider the three following cases: RN black hole representing the solution in asymptotically flat spacetime, Sch-AdS/dS black hole representing the solution in non-asymptotically flat spacetime, and the black hole in the dRGT massive gravity theory representing the solution in modified gravity theories. 
Since the procedure is the same as we have performed, we will list only the important quantities for each solution for conciseness in Appendix~\ref{appen}.

It is worthwhile to note that if we consider the scenario that the nonextensive parameter is held fixed, our results will reduce to those in the studies of black hole thermodynamics found in the literature, e.g., 
the Sch black hole \cite{Biro:2013cra, Czinner:2015eyk, Alonso-Serrano:2020hpb}, 
RN black hole \cite{Promsiri:2022qin},
Kerr black hole \cite{Czinner:2017tjq}, 
Sch-dS black hole \cite{Tannukij:2020njz, Nakarachinda:2021jxd},
and dRGT black hole/string \cite{Sriling:2021lpr, Chunaksorn:2022whl}.
In addition, there are also further studies on the thermodynamic properties of the RN black hole by treating $\lambda$ as a non-fixed thermodynamic variable \cite{Promsiri:2020jga, Promsiri:2021hhv}. 
In these works, the conjugate variable of $\lambda$, which can be interpreted as a volume, is obtained from the leading order term of the small-$\lambda$ approximation as discussed previously.
However, the full structure of the thermodynamic behaviors of black holes has not been investigated yet. 
It is interesting to explore such behaviors.

One can see that it is possible to link the black hole thermodynamics with R\'enyi entropy to the gravitational description. 
Hence, this allows us to suitably investigate the thermodynamic properties of the black holes, e.g., thermodynamic stability and phase transition, as nonextensive systems which are described by the R\'enyi entropy.
In other words, one can properly define other thermodynamic functions such as free energy and heat capacity. 
Besides the extension to other generalized entropies, such as R\'enyi entropy, the efforts for re-establishing a black hole's quantity as the homogeneous function of degree 1 and, consequently, the Smarr formula appears similar to the Euler relation in the standard thermodynamics \cite{Biro:2017flp,Biro:2019rms}.

\section{Conclusion and Discussion}\label{sec: concl}

The connection between classical gravity theory and thermodynamics is deeply based on the area law of the black hole, $S = A/(4G)$.
This provides intensive investigations of the thermodynamic properties of black holes. 
However, the entropy of the black hole is the nonextensive quantity, then the thermodynamic system associated with the black hole should be considered as the nonextensive system. In this article, the form of the nonextensive entropy of the black hole is chosen to be analyzed via the R\'enyi entropy in order to study the thermodynamics of the black hole. R\'enyi entropy does not only describe the equilibrium thermodynamics of a nonextensive system but also satisfies the zeroth law of thermodynamics in which the empirical temperature can be suitably defined.

However, as we reviewed in Sec.~\ref{sec:GB}, the thermodynamic equations of the black hole, e.g., the equation for the first law $\text{d}M= T \text{d}S$ are obtained by identifying the entropy as the BH entropy $S= S_\text{BH} = A/(4G)$ but not as the R\'enyi entropy $S_\text{R} = \frac{1}{\lambda}\ln (1+\lambda S_\text{BH})$. As a result, the equation for the first law of the black hole with R\'enyi entropy is adopted as $\text{d}M= T_\text{R} \text{d}S_\text{R}$ without deriving directly from the gravitational description. This is one of the drawbacks of the investigation of black hole thermodynamics with R\'enyi entropy found in the literature. In this article, we showed that it is possible to overcome such drawbacks by identifying the nonextensive parameter as the thermodynamics variables. By using dimensional analysis, the nonextensive parameter can be identified as the pressure $\lambda = \frac{32}{3}G^2 P_\lambda$ analogous to the role of the cosmological constant in Sch-AdS black hole. As a result, the thermodynamic volume can be identified as a conjugated variable to the pressure via the Legendre structure. Note that the temperature is also obtained as the conjugate variable to the entropy.

One of the important procedures, in order to derive the equation for the first law, is identifying the homogeneous function and then applying Euler's theorem as well as the Smarr formula which is directly obtained from the gravitational description. The procedure is general and can be applied to various kinds of black holes. In this article as an example, we  show that it is possible to apply this procedure to not only the static black hole such as Sch black hole but also the stationary black hole represented by Kerr black hole, non-asymptotically flat black hole represented by Sch-AdS/dS black hole and a black hole in modified gravity theory such as dRGT black hole. It is worthwhile to note that this might be a useful link between gravitational description and black hole thermodynamics with other forms of generalized entropies.

It was pointed out in Ref.~\cite{Nojiri:2021czz} that, by keeping the black hole mass as the ADM mass, the black hole temperature needs to be in the form of the Hawking temperature while using  R\'enyi temperature  the black hole mass is needed to be modified and then the modified mass must satisfy other gravity theories rather than GR.  This issue can be clarified by finding the first law of black hole thermodynamics related to the gravitational description as found in the results of the present work. It is found that the consistency between the gravitational equations and thermodynamic equations is obtained. In fact, the black hole temperature can be suitably interpreted as the R\'enyi temperature while the black hole mass is still the ADM mass.

Besides using the generalized entropy to investigate the thermodynamic properties of the black holes, there have been intensive investigations on the role of generalized entropy in the context of cosmology via the so-called holographic dark energy (HDE) \cite{Tavayef:2018xwx,Moradpour:2018ivi, Saridakis:2020zol,SayahianJahromi:2018irq,Nojiri:2022aof,Nojiri:2022dkr,Nojiri:2022ljp,Nojiri:2022nmu}. 
For the HDE, the energy density of the dark energy is obtained from the notion that  the matter entropy in quantum field theory is not greater than the black hole entropy \cite{Cohen:1998zx}.  Recently, it was argued that the energy density of the dark energy in the HDE model can be obtained from the first law of black hole thermodynamics \cite{Nakarachinda:2022mlz}. While the equation for the first law presented in this paper is taken into account,  it is interesting to investigate the role of generalized entropy such as R\'enyi entropy in the cosmological context. We leave this investigation for further work.

While the quantum field theory is used to study Hawking radiation by incorporating quantum effects on the horizon of the black hole, it is worthwhile to emphasize that our results provide the link between classical gravity and thermodynamics with R\'enyi entropy and do not relate quantum effects on the horizon. Therefore, the R\'enyi temperature cannot be directly derived from the description of quantum field theory to curved spacetime, and then it may not be interpreted as physical quantity as discussed in Ref.~\cite{Nojiri:2021czz}. However, the derivation of the Hawking temperature in Refs.~\cite{Hawking:1975vcx, Hawking:1976ra} seems to adhere to the GB statistics corresponding to the extensive system while the resulting entropy is nonextensive. Therefore, while the entropic nature of a nonextensive system can be debatable, there is no guarantee that the Hawking temperature of a black hole must be derived by using the GB statistics. It is worthwhile to note that there will be more useful if the quantum effects are taken into account by incorporating the nonextensivity. Even though our results are based on classical gravity, it may pave the way to derive the R\'enyi temperature using the notion of quantum field in curved spacetime and we leave this investigation for further works.

\section*{Acknowledgement}
This research project is supported by the Second Century Fund (C2F), Chulalongkorn University, and has also received funding support from the NSRF via the Program Management Unit for Human Resources \& Institutional Development, Research and Innovation [grant number B37G 660013].

\section*{Declarations}
The authors have no conflicts of interest to declare that are relevant to the content of this article.

\section*{Data availability statement}
No new data were created or analyzed in this study.

\begin{appendix}

\section{Important quantities for various black holes described by R\'enyi entropy}\label{appen}
 
As discussed in Secs.~\ref{sec:GB} and \ref{sec: Renyi}, we have expressed the mass of the black holes as a suitable (homogeneous) function of the considered entropy and other quantities. 
Then, by using Euler's theorem of a homogeneous function, one can derive the Smarr formula and further obtain the equation for the first law.
In this section, we will show the explicit expressions of the mass, Smarr formula, first law, and thermodynamic quantities for the RN black hole, Sch-AdS/dS black hole, and dRGT black hole described by the R\'enyi entropy as follows.  
\begin{itemize}
    \item \textbf{RN black hole}
    \begin{itemize}
    \item Black hole mass: 
    \begin{eqnarray}
    M =\frac{\pi \lambda q^2+G(e^{\lambda  S_\text{R}}-1)}{2 \sqrt{\pi G^3\lambda(e^{\lambda  S_R}-1)}}.
    \end{eqnarray}
    \item Homogeneous function degree $1/2$: 
    \begin{eqnarray}
    M(a S_\text{R}, a \lambda^{-1},a q^2 ) = a^{1/2} M( S_\text{R},  \lambda^{-1},q^2 ).
    \end{eqnarray}
    \item Smarr formula: 
    \begin{eqnarray}
	M = 2 S_\text{R} T_\text{R} - 2 \Phi_\lambda\lambda +  \Phi_q q.
    \end{eqnarray}
    \item Equation for the first law: 
    \begin{eqnarray}
    \text{d}M =  T_\text{R} \text{d}S_\text{R} + \Phi_\lambda \text{d}\lambda + \Phi_q \text{d}q.
    \end{eqnarray}
    \item Conjugate variables:
    \begin{eqnarray}
	T_\text{R} &=& \left(\frac{\partial M}{\partial S_\text{R}}\right)_{\lambda, q}
	=\frac{\left(r_H^2-q^2\right) \left(\pi  \lambda  r_H^2+G\right)}{4 \pi G  r_H^3},\\
	\Phi_\lambda &=& \left(\frac{\partial M}{\partial \lambda}\right)_{S_\text{R},q}
	= \frac{\left(r_H^2-q^2\right) \left[\left(G+\pi  \lambda  r_H^2\right) \ln \left(\frac{\pi  \lambda  r_H^2}{G}+1\right)-\pi  \lambda  r_H^2\right]}{4 \pi  G \lambda ^2 r_H^3},\\
	\Phi_q &=& \left(\frac{\partial M}{\partial q}\right)_{S_\text{R},\lambda} = \frac{q}{G r_H}.
    \end{eqnarray}
    \end{itemize}
    \item \textbf{Sch-AdS/dS black hole}
    \begin{itemize}
    \item Black hole mass: 
    \begin{eqnarray}
	M =\frac{1}{6} \sqrt{\frac{e^{\lambda  S_\text{R}}-1}{\pi^3G \lambda^3 }} \Big[3 \pi  \lambda -G \Lambda  \left(e^{\lambda  S_\text{R}}-1\right)\Big].
    \end{eqnarray}
    \item Homogeneous function degree $1/2$: 
    \begin{eqnarray}
	M(a S_\text{R}, a \lambda^{-1}, a \Lambda^{-1}) = a^{1/2} M( S_\text{R},  \lambda^{-1},\Lambda^{-1} ).
    \end{eqnarray}
    \item Smarr formula: 
    \begin{eqnarray}
	M = 2 T_\text{R}S_\text{R} - 2 \Phi_\lambda\lambda -2 \Phi_\Lambda \Lambda.
    \end{eqnarray}
    \item Equation for the first law: 
    \begin{eqnarray}
    	\text{d}M =  T_\text{R} \text{d}S_\text{R} + \Phi_\lambda \text{d}\lambda + \Phi_\Lambda \text{d}\Lambda.
    \end{eqnarray}
    \item Conjugate variables:
    \begin{eqnarray}
    	T_\text{R} &=& \left(\frac{\partial M}{\partial S_R}\right)_{\lambda,\Lambda} 
	=\frac{\left(1-\Lambda  r_H^2\right) \left(G+\pi  \lambda  r_H^2\right)}{4 \pi  G r_H},\\
    	\Phi_\lambda &=& \left(\frac{\partial M}{\partial \lambda}\right)_{S_\text{R},\Lambda} 
	= \frac{\left(1-\Lambda  r_H^2\right) \left[\left(G+\pi  \lambda  r_H^2\right) \ln \left(\frac{\pi  \lambda  r_H^2}{G}+1\right)-\pi  \lambda  r_H^2\right]}{4 \pi  G \lambda ^2 r_H},\\
	\Phi_\Lambda &=& \left(\frac{\partial M}{\partial \Lambda}\right)_{S_\text{R},\lambda} = -\frac{r_H^3}{6 G}.
    \end{eqnarray}
    \end{itemize}
    \item \textbf{dRGT black hole}
    \begin{itemize}
    \item Black hole mass: 
    \begin{eqnarray}
    	M = \sqrt{\frac{e^{\lambda  S_\text{R}}-1}{4\pi^3 G \lambda^3}} \left[m_g^2 \left\{\pi  c_0 \lambda + c_1 \sqrt{\pi G \lambda  \left(e^{\lambda  S_\text{R}}-1\right)}-c_2 G \left(e^{\lambda  S_\text{R}}-1\right)\right\}+\pi  \lambda\right].\,\,\,\,\,
    \end{eqnarray}
    \item Homogeneous function degree $1/2$: 
    \begin{eqnarray}
	M(a S_\text{R}, a \lambda^{-1},a  m_g^{-2},a c_0, a c_1^2) = a^{1/2} M( S_\text{R}, \lambda^{-1},m_g^{-2},c_0, c_1^2 ).
    \end{eqnarray}
    \item Smarr formula: 
    \begin{eqnarray}
     M( S_\text{R},  \lambda^{-1},m_g^{-2},c_0, c_1^2 ) = 2 T_\text{R}S_\text{R} - 2 \Phi_\lambda \lambda -2  \Phi_{m_g} m_g^2+2\Phi_{c_0} c_0+\Phi_{c_1} c_1.
    \end{eqnarray}
    \item Equation for the first law: 
    \begin{eqnarray}
	\text{d}M =  T_\text{R} \text{d}S_\text{R} + \Phi_\lambda \text{d}\lambda + \Phi_{m_g} \text{d} m_g^2 +\Phi_{c_0} \text{d}c_0 +\Phi_{c_1} \text{d}c_1.
    \end{eqnarray}
    \item Conjugate variables:
    \begin{eqnarray}      
    	T_\text{R} &=& \left(\frac{\partial M}{\partial S_R}\right)_{\lambda,m_g,c_0,c_1} 
	=\frac{\left[1- m_g^2 (3  c_2r_H^2-2c_1 r_H - c_0)\right] \left(G+\pi  \lambda  r_H^2\right)}{4 \pi  G r_H},\\
    	\Phi_\lambda &=& \left(\frac{\partial M}{\partial \lambda}\right)_{S_\text{R},m_g,c_0,c_1}\nonumber\\
	&=& \frac{\left[1- m_g^2 (3  c_2r_H^2-2c_1 r_H - c_0)\right]}{4 \pi  G \lambda ^2 r_H}\nonumber\\
	&&\times\left[\left(G+\pi  \lambda  r_H^2\right) \ln \left(\frac{\pi  \lambda  r_H^2}{G}+1\right)-\pi  \lambda  r_H^2\right],\
	\\
	\Phi_{m_g} &=& \left(\frac{\partial M}{\partial m_g^2} \right)_{S_\text{R},\lambda,c_0,c_1}
	= \frac{c_0 r_H +c_1 r_H^2 -c_2 r_H^3}{2 G},\\
	\Phi_{c_0} &=& \left(\frac{\partial M}{\partial c_0} \right)_{S_\text{R},\lambda,m_g,c_1}
	= \frac{m_g^2 r_H }{2 G},\\
	\Phi_{c_1} &=& \left(\frac{\partial M}{\partial c_1} \right)_{S_\text{R},\lambda,m_g,c_0}
	=\frac{m_g^2 r_H^2 }{2 G}.
    \end{eqnarray}
    \end{itemize}
\end{itemize}

\end{appendix}

\bibliography{ref_revised}

\begin{thebibliography}{65}%
\makeatletter
\providecommand \@ifxundefined [1]{%
 \@ifx{#1\undefined}
}%
\providecommand \@ifnum [1]{%
 \ifnum #1\expandafter \@firstoftwo
 \else \expandafter \@secondoftwo
 \fi
}%
\providecommand \@ifx [1]{%
 \ifx #1\expandafter \@firstoftwo
 \else \expandafter \@secondoftwo
 \fi
}%
\providecommand \natexlab [1]{#1}%
\providecommand \enquote  [1]{``#1''}%
\providecommand \bibnamefont  [1]{#1}%
\providecommand \bibfnamefont [1]{#1}%
\providecommand \citenamefont [1]{#1}%
\providecommand \href@noop [0]{\@secondoftwo}%
\providecommand \href [0]{\begingroup \@sanitize@url \@href}%
\providecommand \@href[1]{\@@startlink{#1}\@@href}%
\providecommand \@@href[1]{\endgroup#1\@@endlink}%
\providecommand \@sanitize@url [0]{\catcode `\\12\catcode `\$12\catcode
  `\&12\catcode `\#12\catcode `\^12\catcode `\_12\catcode `\%12\relax}%
\providecommand \@@startlink[1]{}%
\providecommand \@@endlink[0]{}%
\providecommand \url  [0]{\begingroup\@sanitize@url \@url }%
\providecommand \@url [1]{\endgroup\@href {#1}{\urlprefix }}%
\providecommand \urlprefix  [0]{URL }%
\providecommand \Eprint [0]{\href }%
\providecommand \doibase [0]{https://doi.org/}%
\providecommand \selectlanguage [0]{\@gobble}%
\providecommand \bibinfo  [0]{\@secondoftwo}%
\providecommand \bibfield  [0]{\@secondoftwo}%
\providecommand \translation [1]{[#1]}%
\providecommand \BibitemOpen [0]{}%
\providecommand \bibitemStop [0]{}%
\providecommand \bibitemNoStop [0]{.\EOS\space}%
\providecommand \EOS [0]{\spacefactor3000\relax}%
\providecommand \BibitemShut  [1]{\csname bibitem#1\endcsname}%
\let\auto@bib@innerbib\@empty
\bibitem [{\citenamefont {Akiyama}\ \emph
  {et~al.}(2019{\natexlab{a}})\citenamefont {Akiyama} \emph
  {et~al.}}]{EventHorizonTelescope:2019dse}%
  \BibitemOpen
  \bibfield  {author} {\bibinfo {author} {\bibfnamefont {K.}~\bibnamefont
  {Akiyama}} \emph {et~al.} (\bibinfo {collaboration} {Event Horizon
  Telescope}),\ }\bibfield  {title} {\bibinfo {title} {{First M87 Event Horizon
  Telescope Results. I. The Shadow of the Supermassive Black Hole}},\ }\href
  {https://doi.org/10.3847/2041-8213/ab0ec7} {\bibfield  {journal} {\bibinfo
  {journal} {Astrophys. J. Lett.}\ }\textbf {\bibinfo {volume} {875}},\
  \bibinfo {pages} {L1} (\bibinfo {year} {2019}{\natexlab{a}})},\ \Eprint
  {https://arxiv.org/abs/1906.11238} {arXiv:1906.11238 [astro-ph.GA]}
  \BibitemShut {NoStop}%
\bibitem [{\citenamefont {Akiyama}\ \emph
  {et~al.}(2019{\natexlab{b}})\citenamefont {Akiyama} \emph
  {et~al.}}]{EventHorizonTelescope:2019uob}%
  \BibitemOpen
  \bibfield  {author} {\bibinfo {author} {\bibfnamefont {K.}~\bibnamefont
  {Akiyama}} \emph {et~al.} (\bibinfo {collaboration} {Event Horizon
  Telescope}),\ }\bibfield  {title} {\bibinfo {title} {{First M87 Event Horizon
  Telescope Results. II. Array and Instrumentation}},\ }\href
  {https://doi.org/10.3847/2041-8213/ab0c96} {\bibfield  {journal} {\bibinfo
  {journal} {Astrophys. J. Lett.}\ }\textbf {\bibinfo {volume} {875}},\
  \bibinfo {pages} {L2} (\bibinfo {year} {2019}{\natexlab{b}})},\ \Eprint
  {https://arxiv.org/abs/1906.11239} {arXiv:1906.11239 [astro-ph.IM]}
  \BibitemShut {NoStop}%
\bibitem [{\citenamefont {Akiyama}\ \emph
  {et~al.}(2019{\natexlab{c}})\citenamefont {Akiyama} \emph
  {et~al.}}]{EventHorizonTelescope:2019jan}%
  \BibitemOpen
  \bibfield  {author} {\bibinfo {author} {\bibfnamefont {K.}~\bibnamefont
  {Akiyama}} \emph {et~al.} (\bibinfo {collaboration} {Event Horizon
  Telescope}),\ }\bibfield  {title} {\bibinfo {title} {{First M87 Event Horizon
  Telescope Results. III. Data Processing and Calibration}},\ }\href
  {https://doi.org/10.3847/2041-8213/ab0c57} {\bibfield  {journal} {\bibinfo
  {journal} {Astrophys. J. Lett.}\ }\textbf {\bibinfo {volume} {875}},\
  \bibinfo {pages} {L3} (\bibinfo {year} {2019}{\natexlab{c}})},\ \Eprint
  {https://arxiv.org/abs/1906.11240} {arXiv:1906.11240 [astro-ph.GA]}
  \BibitemShut {NoStop}%
\bibitem [{\citenamefont {Akiyama}\ \emph
  {et~al.}(2019{\natexlab{d}})\citenamefont {Akiyama} \emph
  {et~al.}}]{EventHorizonTelescope:2019ths}%
  \BibitemOpen
  \bibfield  {author} {\bibinfo {author} {\bibfnamefont {K.}~\bibnamefont
  {Akiyama}} \emph {et~al.} (\bibinfo {collaboration} {Event Horizon
  Telescope}),\ }\bibfield  {title} {\bibinfo {title} {{First M87 Event Horizon
  Telescope Results. IV. Imaging the Central Supermassive Black Hole}},\ }\href
  {https://doi.org/10.3847/2041-8213/ab0e85} {\bibfield  {journal} {\bibinfo
  {journal} {Astrophys. J. Lett.}\ }\textbf {\bibinfo {volume} {875}},\
  \bibinfo {pages} {L4} (\bibinfo {year} {2019}{\natexlab{d}})},\ \Eprint
  {https://arxiv.org/abs/1906.11241} {arXiv:1906.11241 [astro-ph.GA]}
  \BibitemShut {NoStop}%
\bibitem [{\citenamefont {Akiyama}\ \emph
  {et~al.}(2019{\natexlab{e}})\citenamefont {Akiyama} \emph
  {et~al.}}]{EventHorizonTelescope:2019pgp}%
  \BibitemOpen
  \bibfield  {author} {\bibinfo {author} {\bibfnamefont {K.}~\bibnamefont
  {Akiyama}} \emph {et~al.} (\bibinfo {collaboration} {Event Horizon
  Telescope}),\ }\bibfield  {title} {\bibinfo {title} {{First M87 Event Horizon
  Telescope Results. V. Physical Origin of the Asymmetric Ring}},\ }\href
  {https://doi.org/10.3847/2041-8213/ab0f43} {\bibfield  {journal} {\bibinfo
  {journal} {Astrophys. J. Lett.}\ }\textbf {\bibinfo {volume} {875}},\
  \bibinfo {pages} {L5} (\bibinfo {year} {2019}{\natexlab{e}})},\ \Eprint
  {https://arxiv.org/abs/1906.11242} {arXiv:1906.11242 [astro-ph.GA]}
  \BibitemShut {NoStop}%
\bibitem [{\citenamefont {Akiyama}\ \emph
  {et~al.}(2019{\natexlab{f}})\citenamefont {Akiyama} \emph
  {et~al.}}]{EventHorizonTelescope:2019ggy}%
  \BibitemOpen
  \bibfield  {author} {\bibinfo {author} {\bibfnamefont {K.}~\bibnamefont
  {Akiyama}} \emph {et~al.} (\bibinfo {collaboration} {Event Horizon
  Telescope}),\ }\bibfield  {title} {\bibinfo {title} {{First M87 Event Horizon
  Telescope Results. VI. The Shadow and Mass of the Central Black Hole}},\
  }\href {https://doi.org/10.3847/2041-8213/ab1141} {\bibfield  {journal}
  {\bibinfo  {journal} {Astrophys. J. Lett.}\ }\textbf {\bibinfo {volume}
  {875}},\ \bibinfo {pages} {L6} (\bibinfo {year} {2019}{\natexlab{f}})},\
  \Eprint {https://arxiv.org/abs/1906.11243} {arXiv:1906.11243 [astro-ph.GA]}
  \BibitemShut {NoStop}%
\bibitem [{\citenamefont {Hawking}(1975)}]{Hawking:1975vcx}%
  \BibitemOpen
  \bibfield  {author} {\bibinfo {author} {\bibfnamefont {S.~W.}\ \bibnamefont
  {Hawking}},\ }\bibfield  {title} {\bibinfo {title} {{Particle Creation by
  Black Holes}},\ }\href {https://doi.org/10.1007/BF02345020} {\bibfield
  {journal} {\bibinfo  {journal} {Commun. Math. Phys.}\ }\textbf {\bibinfo
  {volume} {43}},\ \bibinfo {pages} {199} (\bibinfo {year} {1975})},\ \bibinfo
  {note} {[Erratum: Commun.Math.Phys. 46, 206 (1976)]}\BibitemShut {NoStop}%
\bibitem [{\citenamefont {Hawking}(1971)}]{Hawking:1971tu}%
  \BibitemOpen
  \bibfield  {author} {\bibinfo {author} {\bibfnamefont {S.~W.}\ \bibnamefont
  {Hawking}},\ }\bibfield  {title} {\bibinfo {title} {{Gravitational radiation
  from colliding black holes}},\ }\href
  {https://doi.org/10.1103/PhysRevLett.26.1344} {\bibfield  {journal} {\bibinfo
   {journal} {Phys. Rev. Lett.}\ }\textbf {\bibinfo {volume} {26}},\ \bibinfo
  {pages} {1344} (\bibinfo {year} {1971})}\BibitemShut {NoStop}%
\bibitem [{\citenamefont {Bekenstein}(1973)}]{Bekenstein:1973ur}%
  \BibitemOpen
  \bibfield  {author} {\bibinfo {author} {\bibfnamefont {J.~D.}\ \bibnamefont
  {Bekenstein}},\ }\bibfield  {title} {\bibinfo {title} {{Black holes and
  entropy}},\ }\href {https://doi.org/10.1103/PhysRevD.7.2333} {\bibfield
  {journal} {\bibinfo  {journal} {Phys. Rev. D}\ }\textbf {\bibinfo {volume}
  {7}},\ \bibinfo {pages} {2333} (\bibinfo {year} {1973})}\BibitemShut
  {NoStop}%
\bibitem [{\citenamefont {Bardeen}\ \emph {et~al.}(1973)\citenamefont
  {Bardeen}, \citenamefont {Carter},\ and\ \citenamefont
  {Hawking}}]{Bardeen:1973gs}%
  \BibitemOpen
  \bibfield  {author} {\bibinfo {author} {\bibfnamefont {J.~M.}\ \bibnamefont
  {Bardeen}}, \bibinfo {author} {\bibfnamefont {B.}~\bibnamefont {Carter}},\
  and\ \bibinfo {author} {\bibfnamefont {S.~W.}\ \bibnamefont {Hawking}},\
  }\bibfield  {title} {\bibinfo {title} {{The Four laws of black hole
  mechanics}},\ }\href {https://doi.org/10.1007/BF01645742} {\bibfield
  {journal} {\bibinfo  {journal} {Commun. Math. Phys.}\ }\textbf {\bibinfo
  {volume} {31}},\ \bibinfo {pages} {161} (\bibinfo {year} {1973})}\BibitemShut
  {NoStop}%
\bibitem [{\citenamefont {R{\'e}nyi}(1959)}]{Renyi1959}%
  \BibitemOpen
  \bibfield  {author} {\bibinfo {author} {\bibfnamefont {A.}~\bibnamefont
  {R{\'e}nyi}},\ }\bibfield  {title} {\bibinfo {title} {On the dimension and
  entropy of probability distributions},\ }\href
  {https://doi.org/10.1007/BF02063299} {\bibfield  {journal} {\bibinfo
  {journal} {Acta Mathematica Academiae Scientiarum Hungarica}\ }\textbf
  {\bibinfo {volume} {10}},\ \bibinfo {pages} {193} (\bibinfo {year}
  {1959})}\BibitemShut {NoStop}%
\bibitem [{\citenamefont {\c{C}imdiker}\ \emph {et~al.}(2023)\citenamefont
  {\c{C}imdiker}, \citenamefont {D\c{a}browski},\ and\ \citenamefont
  {Gohar}}]{Cimdiker:2022ics}%
  \BibitemOpen
  \bibfield  {author} {\bibinfo {author} {\bibfnamefont {I.}~\bibnamefont
  {\c{C}imdiker}}, \bibinfo {author} {\bibfnamefont {M.~P.}\ \bibnamefont
  {D\c{a}browski}},\ and\ \bibinfo {author} {\bibfnamefont {H.}~\bibnamefont
  {Gohar}},\ }\bibfield  {title} {\bibinfo {title} {{Equilibrium temperature
  for black holes with nonextensive entropy}},\ }\href
  {https://doi.org/10.1140/epjc/s10052-023-11317-0} {\bibfield  {journal}
  {\bibinfo  {journal} {Eur. Phys. J. C}\ }\textbf {\bibinfo {volume} {83}},\
  \bibinfo {pages} {169} (\bibinfo {year} {2023})},\ \Eprint
  {https://arxiv.org/abs/2208.04473} {arXiv:2208.04473 [gr-qc]} \BibitemShut
  {NoStop}%
\bibitem [{\citenamefont {Bir\'o}\ and\ \citenamefont
  {Czinner}(2013)}]{Biro:2013cra}%
  \BibitemOpen
  \bibfield  {author} {\bibinfo {author} {\bibfnamefont {T.~S.}\ \bibnamefont
  {Bir\'o}}\ and\ \bibinfo {author} {\bibfnamefont {V.~G.}\ \bibnamefont
  {Czinner}},\ }\bibfield  {title} {\bibinfo {title} {{A $q$-parameter bound
  for particle spectra based on black hole thermodynamics with R\'enyi
  entropy}},\ }\href {https://doi.org/10.1016/j.physletb.2013.09.032}
  {\bibfield  {journal} {\bibinfo  {journal} {Phys. Lett. B}\ }\textbf
  {\bibinfo {volume} {726}},\ \bibinfo {pages} {861} (\bibinfo {year}
  {2013})},\ \Eprint {https://arxiv.org/abs/1309.4261} {arXiv:1309.4261
  [gr-qc]} \BibitemShut {NoStop}%
\bibitem [{\citenamefont {Czinner}\ and\ \citenamefont
  {Iguchi}(2016)}]{Czinner:2015eyk}%
  \BibitemOpen
  \bibfield  {author} {\bibinfo {author} {\bibfnamefont {V.~G.}\ \bibnamefont
  {Czinner}}\ and\ \bibinfo {author} {\bibfnamefont {H.}~\bibnamefont
  {Iguchi}},\ }\bibfield  {title} {\bibinfo {title} {{R\'enyi Entropy and the
  Thermodynamic Stability of Black Holes}},\ }\href
  {https://doi.org/10.1016/j.physletb.2015.11.061} {\bibfield  {journal}
  {\bibinfo  {journal} {Phys. Lett. B}\ }\textbf {\bibinfo {volume} {752}},\
  \bibinfo {pages} {306} (\bibinfo {year} {2016})},\ \Eprint
  {https://arxiv.org/abs/1511.06963} {arXiv:1511.06963 [gr-qc]} \BibitemShut
  {NoStop}%
\bibitem [{\citenamefont {Czinner}\ and\ \citenamefont
  {Iguchi}(2017)}]{Czinner:2017tjq}%
  \BibitemOpen
  \bibfield  {author} {\bibinfo {author} {\bibfnamefont {V.~G.}\ \bibnamefont
  {Czinner}}\ and\ \bibinfo {author} {\bibfnamefont {H.}~\bibnamefont
  {Iguchi}},\ }\bibfield  {title} {\bibinfo {title} {{Thermodynamics, stability
  and Hawking\textendash{}Page transition of Kerr black holes from R\'enyi
  statistics}},\ }\href {https://doi.org/10.1140/epjc/s10052-017-5453-x}
  {\bibfield  {journal} {\bibinfo  {journal} {Eur. Phys. J. C}\ }\textbf
  {\bibinfo {volume} {77}},\ \bibinfo {pages} {892} (\bibinfo {year} {2017})},\
  \Eprint {https://arxiv.org/abs/1702.05341} {arXiv:1702.05341 [gr-qc]}
  \BibitemShut {NoStop}%
\bibitem [{\citenamefont {Promsiri}\ \emph {et~al.}(2021)\citenamefont
  {Promsiri}, \citenamefont {Hirunsirisawat},\ and\ \citenamefont
  {Liewrian}}]{Promsiri:2021hhv}%
  \BibitemOpen
  \bibfield  {author} {\bibinfo {author} {\bibfnamefont {C.}~\bibnamefont
  {Promsiri}}, \bibinfo {author} {\bibfnamefont {E.}~\bibnamefont
  {Hirunsirisawat}},\ and\ \bibinfo {author} {\bibfnamefont {W.}~\bibnamefont
  {Liewrian}},\ }\bibfield  {title} {\bibinfo {title} {{Solid-liquid phase
  transition and heat engine in an asymptotically flat Schwarzschild black hole
  via the R\'enyi extended phase space approach}},\ }\href
  {https://doi.org/10.1103/PhysRevD.104.064004} {\bibfield  {journal} {\bibinfo
   {journal} {Phys. Rev. D}\ }\textbf {\bibinfo {volume} {104}},\ \bibinfo
  {pages} {064004} (\bibinfo {year} {2021})},\ \Eprint
  {https://arxiv.org/abs/2106.02406} {arXiv:2106.02406 [hep-th]} \BibitemShut
  {NoStop}%
\bibitem [{\citenamefont {Alonso-Serrano}\ \emph {et~al.}(2021)\citenamefont
  {Alonso-Serrano}, \citenamefont {Dabrowski},\ and\ \citenamefont
  {Gohar}}]{Alonso-Serrano:2020hpb}%
  \BibitemOpen
  \bibfield  {author} {\bibinfo {author} {\bibfnamefont {A.}~\bibnamefont
  {Alonso-Serrano}}, \bibinfo {author} {\bibfnamefont {M.~P.}\ \bibnamefont
  {Dabrowski}},\ and\ \bibinfo {author} {\bibfnamefont {H.}~\bibnamefont
  {Gohar}},\ }\bibfield  {title} {\bibinfo {title} {{Nonextensive Black Hole
  Entropy and Quantum Gravity Effects at the Last Stages of Evaporation}},\
  }\href {https://doi.org/10.1103/PhysRevD.103.026021} {\bibfield  {journal}
  {\bibinfo  {journal} {Phys. Rev. D}\ }\textbf {\bibinfo {volume} {103}},\
  \bibinfo {pages} {026021} (\bibinfo {year} {2021})},\ \Eprint
  {https://arxiv.org/abs/2009.02129} {arXiv:2009.02129 [gr-qc]} \BibitemShut
  {NoStop}%
\bibitem [{\citenamefont {Promsiri}\ \emph {et~al.}(2020)\citenamefont
  {Promsiri}, \citenamefont {Hirunsirisawat},\ and\ \citenamefont
  {Liewrian}}]{Promsiri:2020jga}%
  \BibitemOpen
  \bibfield  {author} {\bibinfo {author} {\bibfnamefont {C.}~\bibnamefont
  {Promsiri}}, \bibinfo {author} {\bibfnamefont {E.}~\bibnamefont
  {Hirunsirisawat}},\ and\ \bibinfo {author} {\bibfnamefont {W.}~\bibnamefont
  {Liewrian}},\ }\bibfield  {title} {\bibinfo {title} {{Thermodynamics and Van
  der Waals phase transition of charged black holes in flat spacetime via
  R\'enyi statistics}},\ }\href {https://doi.org/10.1103/PhysRevD.102.064014}
  {\bibfield  {journal} {\bibinfo  {journal} {Phys. Rev. D}\ }\textbf {\bibinfo
  {volume} {102}},\ \bibinfo {pages} {064014} (\bibinfo {year} {2020})},\
  \Eprint {https://arxiv.org/abs/2003.12986} {arXiv:2003.12986 [hep-th]}
  \BibitemShut {NoStop}%
\bibitem [{\citenamefont {Tannukij}\ \emph {et~al.}(2020)\citenamefont
  {Tannukij}, \citenamefont {Wongjun}, \citenamefont {Hirunsirisawat},
  \citenamefont {Deesuwan},\ and\ \citenamefont {Promsiri}}]{Tannukij:2020njz}%
  \BibitemOpen
  \bibfield  {author} {\bibinfo {author} {\bibfnamefont {L.}~\bibnamefont
  {Tannukij}}, \bibinfo {author} {\bibfnamefont {P.}~\bibnamefont {Wongjun}},
  \bibinfo {author} {\bibfnamefont {E.}~\bibnamefont {Hirunsirisawat}},
  \bibinfo {author} {\bibfnamefont {T.}~\bibnamefont {Deesuwan}},\ and\
  \bibinfo {author} {\bibfnamefont {C.}~\bibnamefont {Promsiri}},\ }\bibfield
  {title} {\bibinfo {title} {{Thermodynamics and phase transition of
  spherically symmetric black hole in de Sitter space from R\'enyi
  statistics}},\ }\href {https://doi.org/10.1140/epjp/s13360-020-00517-2}
  {\bibfield  {journal} {\bibinfo  {journal} {Eur. Phys. J. Plus}\ }\textbf
  {\bibinfo {volume} {135}},\ \bibinfo {pages} {500} (\bibinfo {year}
  {2020})},\ \Eprint {https://arxiv.org/abs/2002.00377} {arXiv:2002.00377
  [gr-qc]} \BibitemShut {NoStop}%
\bibitem [{\citenamefont {Nakarachinda}\ \emph {et~al.}(2021)\citenamefont
  {Nakarachinda}, \citenamefont {Hirunsirisawat}, \citenamefont {Tannukij},\
  and\ \citenamefont {Wongjun}}]{Nakarachinda:2021jxd}%
  \BibitemOpen
  \bibfield  {author} {\bibinfo {author} {\bibfnamefont {R.}~\bibnamefont
  {Nakarachinda}}, \bibinfo {author} {\bibfnamefont {E.}~\bibnamefont
  {Hirunsirisawat}}, \bibinfo {author} {\bibfnamefont {L.}~\bibnamefont
  {Tannukij}},\ and\ \bibinfo {author} {\bibfnamefont {P.}~\bibnamefont
  {Wongjun}},\ }\bibfield  {title} {\bibinfo {title} {{Effective
  thermodynamical system of Schwarzschild\textendash{}de Sitter black holes
  from R\'enyi statistics}},\ }\href
  {https://doi.org/10.1103/PhysRevD.104.064003} {\bibfield  {journal} {\bibinfo
   {journal} {Phys. Rev. D}\ }\textbf {\bibinfo {volume} {104}},\ \bibinfo
  {pages} {064003} (\bibinfo {year} {2021})},\ \Eprint
  {https://arxiv.org/abs/2106.02838} {arXiv:2106.02838 [gr-qc]} \BibitemShut
  {NoStop}%
\bibitem [{\citenamefont {Sriling}\ \emph {et~al.}(2022)\citenamefont
  {Sriling}, \citenamefont {Nakarachinda},\ and\ \citenamefont
  {Wongjun}}]{Sriling:2021lpr}%
  \BibitemOpen
  \bibfield  {author} {\bibinfo {author} {\bibfnamefont {P.}~\bibnamefont
  {Sriling}}, \bibinfo {author} {\bibfnamefont {R.}~\bibnamefont
  {Nakarachinda}},\ and\ \bibinfo {author} {\bibfnamefont {P.}~\bibnamefont
  {Wongjun}},\ }\bibfield  {title} {\bibinfo {title} {{Thermodynamics of black
  string from R\'enyi entropy in de
  Rham\textendash{}Gabadadze\textendash{}Tolley massive gravity theory}},\
  }\href {https://doi.org/10.1088/1361-6382/ac750b} {\bibfield  {journal}
  {\bibinfo  {journal} {Class. Quant. Grav.}\ }\textbf {\bibinfo {volume}
  {39}},\ \bibinfo {pages} {185006} (\bibinfo {year} {2022})},\ \Eprint
  {https://arxiv.org/abs/2112.13120} {arXiv:2112.13120 [gr-qc]} \BibitemShut
  {NoStop}%
\bibitem [{\citenamefont {Promsiri}\ \emph {et~al.}(2022)\citenamefont
  {Promsiri}, \citenamefont {Hirunsirisawat},\ and\ \citenamefont
  {Nakarachinda}}]{Promsiri:2022qin}%
  \BibitemOpen
  \bibfield  {author} {\bibinfo {author} {\bibfnamefont {C.}~\bibnamefont
  {Promsiri}}, \bibinfo {author} {\bibfnamefont {E.}~\bibnamefont
  {Hirunsirisawat}},\ and\ \bibinfo {author} {\bibfnamefont {R.}~\bibnamefont
  {Nakarachinda}},\ }\bibfield  {title} {\bibinfo {title} {{Emergent phase,
  thermodynamic geometry, and criticality of charged black holes from R\'enyi
  statistics}},\ }\href {https://doi.org/10.1103/PhysRevD.105.124049}
  {\bibfield  {journal} {\bibinfo  {journal} {Phys. Rev. D}\ }\textbf {\bibinfo
  {volume} {105}},\ \bibinfo {pages} {124049} (\bibinfo {year} {2022})},\
  \Eprint {https://arxiv.org/abs/2204.13023} {arXiv:2204.13023 [hep-th]}
  \BibitemShut {NoStop}%
\bibitem [{\citenamefont {Chunaksorn}\ \emph {et~al.}(2022)\citenamefont
  {Chunaksorn}, \citenamefont {Hirunsirisawat}, \citenamefont {Nakarachinda},
  \citenamefont {Tannukij},\ and\ \citenamefont
  {Wongjun}}]{Chunaksorn:2022whl}%
  \BibitemOpen
  \bibfield  {author} {\bibinfo {author} {\bibfnamefont {P.}~\bibnamefont
  {Chunaksorn}}, \bibinfo {author} {\bibfnamefont {E.}~\bibnamefont
  {Hirunsirisawat}}, \bibinfo {author} {\bibfnamefont {R.}~\bibnamefont
  {Nakarachinda}}, \bibinfo {author} {\bibfnamefont {L.}~\bibnamefont
  {Tannukij}},\ and\ \bibinfo {author} {\bibfnamefont {P.}~\bibnamefont
  {Wongjun}},\ }\bibfield  {title} {\bibinfo {title} {{Thermodynamics of
  asymptotically de Sitter black hole in dRGT massive gravity from R\'enyi
  entropy}},\ }\href {https://doi.org/10.1140/epjc/s10052-022-11110-5}
  {\bibfield  {journal} {\bibinfo  {journal} {Eur. Phys. J. C}\ }\textbf
  {\bibinfo {volume} {82}},\ \bibinfo {pages} {1174} (\bibinfo {year}
  {2022})},\ \Eprint {https://arxiv.org/abs/2208.14770} {arXiv:2208.14770
  [gr-qc]} \BibitemShut {NoStop}%
\bibitem [{\citenamefont {Nojiri}\ \emph {et~al.}(2021)\citenamefont {Nojiri},
  \citenamefont {Odintsov},\ and\ \citenamefont {Faraoni}}]{Nojiri:2021czz}%
  \BibitemOpen
  \bibfield  {author} {\bibinfo {author} {\bibfnamefont {S.}~\bibnamefont
  {Nojiri}}, \bibinfo {author} {\bibfnamefont {S.~D.}\ \bibnamefont
  {Odintsov}},\ and\ \bibinfo {author} {\bibfnamefont {V.}~\bibnamefont
  {Faraoni}},\ }\bibfield  {title} {\bibinfo {title} {{Area-law versus R\'enyi
  and Tsallis black hole entropies}},\ }\href
  {https://doi.org/10.1103/PhysRevD.104.084030} {\bibfield  {journal} {\bibinfo
   {journal} {Phys. Rev. D}\ }\textbf {\bibinfo {volume} {104}},\ \bibinfo
  {pages} {084030} (\bibinfo {year} {2021})},\ \Eprint
  {https://arxiv.org/abs/2109.05315} {arXiv:2109.05315 [gr-qc]} \BibitemShut
  {NoStop}%
\bibitem [{\citenamefont {Nojiri}\ \emph
  {et~al.}(2022{\natexlab{a}})\citenamefont {Nojiri}, \citenamefont
  {Odintsov},\ and\ \citenamefont {Faraoni}}]{Nojiri:2022sfd}%
  \BibitemOpen
  \bibfield  {author} {\bibinfo {author} {\bibfnamefont {S.}~\bibnamefont
  {Nojiri}}, \bibinfo {author} {\bibfnamefont {S.~D.}\ \bibnamefont
  {Odintsov}},\ and\ \bibinfo {author} {\bibfnamefont {V.}~\bibnamefont
  {Faraoni}},\ }\bibfield  {title} {\bibinfo {title} {{Alternative entropies
  and consistent black hole thermodynamics}},\ }\href
  {https://doi.org/10.1142/S0219887822502103} {\bibfield  {journal} {\bibinfo
  {journal} {Int. J. Geom. Meth. Mod. Phys.}\ }\textbf {\bibinfo {volume}
  {19}},\ \bibinfo {pages} {2250210} (\bibinfo {year} {2022}{\natexlab{a}})},\
  \Eprint {https://arxiv.org/abs/2207.07905} {arXiv:2207.07905 [gr-qc]}
  \BibitemShut {NoStop}%
\bibitem [{\citenamefont {Komar}(1959)}]{Komar:1958wp}%
  \BibitemOpen
  \bibfield  {author} {\bibinfo {author} {\bibfnamefont {A.}~\bibnamefont
  {Komar}},\ }\bibfield  {title} {\bibinfo {title} {{Covariant conservation
  laws in general relativity}},\ }\href
  {https://doi.org/10.1103/PhysRev.113.934} {\bibfield  {journal} {\bibinfo
  {journal} {Phys. Rev.}\ }\textbf {\bibinfo {volume} {113}},\ \bibinfo {pages}
  {934} (\bibinfo {year} {1959})}\BibitemShut {NoStop}%
\bibitem [{\citenamefont {Katz}(1985)}]{JKatz1985}%
  \BibitemOpen
  \bibfield  {author} {\bibinfo {author} {\bibfnamefont {J.}~\bibnamefont
  {Katz}},\ }\bibfield  {title} {\bibinfo {title} {A note on komar's anomalous
  factor},\ }\href {https://doi.org/10.1088/0264-9381/2/3/018} {\bibfield
  {journal} {\bibinfo  {journal} {Classical and Quantum Gravity}\ }\textbf
  {\bibinfo {volume} {2}},\ \bibinfo {pages} {423} (\bibinfo {year}
  {1985})}\BibitemShut {NoStop}%
\bibitem [{\citenamefont {Poisson}(2009)}]{Poisson:2009pwt}%
  \BibitemOpen
  \bibfield  {author} {\bibinfo {author} {\bibfnamefont {E.}~\bibnamefont
  {Poisson}},\ }\href {https://doi.org/10.1017/CBO9780511606601} {\emph
  {\bibinfo {title} {{A Relativist's Toolkit: The Mathematics of Black-Hole
  Mechanics}}}}\ (\bibinfo  {publisher} {Cambridge University Press},\ \bibinfo
  {year} {2009})\BibitemShut {NoStop}%
\bibitem [{\citenamefont {Padmanabhan}(2014)}]{Padmanabhan:2010zzb}%
  \BibitemOpen
  \bibfield  {author} {\bibinfo {author} {\bibfnamefont {T.}~\bibnamefont
  {Padmanabhan}},\ }\href@noop {} {\emph {\bibinfo {title} {{Gravitation:
  Foundations and frontiers}}}}\ (\bibinfo  {publisher} {Cambridge University
  Press},\ \bibinfo {year} {2014})\BibitemShut {NoStop}%
\bibitem [{\citenamefont {Arnowitt}\ \emph {et~al.}(2008)\citenamefont
  {Arnowitt}, \citenamefont {Deser},\ and\ \citenamefont
  {Misner}}]{Arnowitt:1962hi}%
  \BibitemOpen
  \bibfield  {author} {\bibinfo {author} {\bibfnamefont {R.~L.}\ \bibnamefont
  {Arnowitt}}, \bibinfo {author} {\bibfnamefont {S.}~\bibnamefont {Deser}},\
  and\ \bibinfo {author} {\bibfnamefont {C.~W.}\ \bibnamefont {Misner}},\
  }\bibfield  {title} {\bibinfo {title} {{The Dynamics of general
  relativity}},\ }\href {https://doi.org/10.1007/s10714-008-0661-1} {\bibfield
  {journal} {\bibinfo  {journal} {Gen. Rel. Grav.}\ }\textbf {\bibinfo {volume}
  {40}},\ \bibinfo {pages} {1997} (\bibinfo {year} {2008})},\ \Eprint
  {https://arxiv.org/abs/gr-qc/0405109} {arXiv:gr-qc/0405109} \BibitemShut
  {NoStop}%
\bibitem [{\citenamefont {Smarr}(1973)}]{Smarr:1972kt}%
  \BibitemOpen
  \bibfield  {author} {\bibinfo {author} {\bibfnamefont {L.}~\bibnamefont
  {Smarr}},\ }\bibfield  {title} {\bibinfo {title} {{Mass formula for Kerr
  black holes}},\ }\href {https://doi.org/10.1103/PhysRevLett.30.71} {\bibfield
   {journal} {\bibinfo  {journal} {Phys. Rev. Lett.}\ }\textbf {\bibinfo
  {volume} {30}},\ \bibinfo {pages} {71} (\bibinfo {year} {1973})},\ \bibinfo
  {note} {[Erratum: Phys. Rev. Lett. {\bf30}, 521 (1973)]}\BibitemShut
  {NoStop}%
\bibitem [{\citenamefont {Hawking}\ and\ \citenamefont
  {Ellis}(2011)}]{Hawking:1973uf}%
  \BibitemOpen
  \bibfield  {author} {\bibinfo {author} {\bibfnamefont {S.~W.}\ \bibnamefont
  {Hawking}}\ and\ \bibinfo {author} {\bibfnamefont {G.~F.~R.}\ \bibnamefont
  {Ellis}},\ }\href {https://doi.org/10.1017/CBO9780511524646} {\emph {\bibinfo
  {title} {{The Large Scale Structure of Space-Time}}}},\ Cambridge Monographs
  on Mathematical Physics\ (\bibinfo  {publisher} {Cambridge University
  Press},\ \bibinfo {year} {2011})\BibitemShut {NoStop}%
\bibitem [{\citenamefont {Magnon}(1985)}]{Magnon:1985sc}%
  \BibitemOpen
  \bibfield  {author} {\bibinfo {author} {\bibfnamefont {A.}~\bibnamefont
  {Magnon}},\ }\bibfield  {title} {\bibinfo {title} {{On Komar integrals in
  asymptotically anti-de Sitter space-times}},\ }\href
  {https://doi.org/10.1063/1.526690} {\bibfield  {journal} {\bibinfo  {journal}
  {J. Math. Phys.}\ }\textbf {\bibinfo {volume} {26}},\ \bibinfo {pages} {3112}
  (\bibinfo {year} {1985})}\BibitemShut {NoStop}%
\bibitem [{\citenamefont {Kastor}(2008)}]{Kastor:2008xb}%
  \BibitemOpen
  \bibfield  {author} {\bibinfo {author} {\bibfnamefont {D.}~\bibnamefont
  {Kastor}},\ }\bibfield  {title} {\bibinfo {title} {{Komar Integrals in Higher
  (and Lower) Derivative Gravity}},\ }\href
  {https://doi.org/10.1088/0264-9381/25/17/175007} {\bibfield  {journal}
  {\bibinfo  {journal} {Class. Quant. Grav.}\ }\textbf {\bibinfo {volume}
  {25}},\ \bibinfo {pages} {175007} (\bibinfo {year} {2008})},\ \Eprint
  {https://arxiv.org/abs/0804.1832} {arXiv:0804.1832 [hep-th]} \BibitemShut
  {NoStop}%
\bibitem [{\citenamefont {Kastor}\ \emph {et~al.}(2009)\citenamefont {Kastor},
  \citenamefont {Ray},\ and\ \citenamefont {Traschen}}]{Kastor:2009wy}%
  \BibitemOpen
  \bibfield  {author} {\bibinfo {author} {\bibfnamefont {D.}~\bibnamefont
  {Kastor}}, \bibinfo {author} {\bibfnamefont {S.}~\bibnamefont {Ray}},\ and\
  \bibinfo {author} {\bibfnamefont {J.}~\bibnamefont {Traschen}},\ }\bibfield
  {title} {\bibinfo {title} {{Enthalpy and the Mechanics of AdS Black Holes}},\
  }\href {https://doi.org/10.1088/0264-9381/26/19/195011} {\bibfield  {journal}
  {\bibinfo  {journal} {Class. Quant. Grav.}\ }\textbf {\bibinfo {volume}
  {26}},\ \bibinfo {pages} {195011} (\bibinfo {year} {2009})},\ \Eprint
  {https://arxiv.org/abs/0904.2765} {arXiv:0904.2765 [hep-th]} \BibitemShut
  {NoStop}%
\bibitem [{\citenamefont {Caldarelli}\ \emph {et~al.}(2000)\citenamefont
  {Caldarelli}, \citenamefont {Cognola},\ and\ \citenamefont
  {Klemm}}]{Caldarelli:1999xj}%
  \BibitemOpen
  \bibfield  {author} {\bibinfo {author} {\bibfnamefont {M.~M.}\ \bibnamefont
  {Caldarelli}}, \bibinfo {author} {\bibfnamefont {G.}~\bibnamefont
  {Cognola}},\ and\ \bibinfo {author} {\bibfnamefont {D.}~\bibnamefont
  {Klemm}},\ }\bibfield  {title} {\bibinfo {title} {{Thermodynamics of
  Kerr-Newman-AdS black holes and conformal field theories}},\ }\href
  {https://doi.org/10.1088/0264-9381/17/2/310} {\bibfield  {journal} {\bibinfo
  {journal} {Class. Quant. Grav.}\ }\textbf {\bibinfo {volume} {17}},\ \bibinfo
  {pages} {399} (\bibinfo {year} {2000})},\ \Eprint
  {https://arxiv.org/abs/hep-th/9908022} {arXiv:hep-th/9908022} \BibitemShut
  {NoStop}%
\bibitem [{\citenamefont {Awad}\ and\ \citenamefont
  {Johnson}(2000)}]{Awad:1999xx}%
  \BibitemOpen
  \bibfield  {author} {\bibinfo {author} {\bibfnamefont {A.~M.}\ \bibnamefont
  {Awad}}\ and\ \bibinfo {author} {\bibfnamefont {C.~V.}\ \bibnamefont
  {Johnson}},\ }\bibfield  {title} {\bibinfo {title} {{Holographic stress
  tensors for Kerr - AdS black holes}},\ }\href
  {https://doi.org/10.1103/PhysRevD.61.084025} {\bibfield  {journal} {\bibinfo
  {journal} {Phys. Rev. D}\ }\textbf {\bibinfo {volume} {61}},\ \bibinfo
  {pages} {084025} (\bibinfo {year} {2000})},\ \Eprint
  {https://arxiv.org/abs/hep-th/9910040} {arXiv:hep-th/9910040} \BibitemShut
  {NoStop}%
\bibitem [{\citenamefont {Das}\ and\ \citenamefont {Mann}(2000)}]{Das:2000cu}%
  \BibitemOpen
  \bibfield  {author} {\bibinfo {author} {\bibfnamefont {S.}~\bibnamefont
  {Das}}\ and\ \bibinfo {author} {\bibfnamefont {R.~B.}\ \bibnamefont {Mann}},\
  }\bibfield  {title} {\bibinfo {title} {{Conserved quantities in Kerr-anti-de
  Sitter space-times in various dimensions}},\ }\href
  {https://doi.org/10.1088/1126-6708/2000/08/033} {\bibfield  {journal}
  {\bibinfo  {journal} {JHEP}\ }\textbf {\bibinfo {volume} {08}},\ \bibinfo
  {pages} {033}},\ \Eprint {https://arxiv.org/abs/hep-th/0008028}
  {arXiv:hep-th/0008028} \BibitemShut {NoStop}%
\bibitem [{\citenamefont {Hajian}\ and\ \citenamefont
  {Sheikh-Jabbari}(2016)}]{Hajian:2015xlp}%
  \BibitemOpen
  \bibfield  {author} {\bibinfo {author} {\bibfnamefont {K.}~\bibnamefont
  {Hajian}}\ and\ \bibinfo {author} {\bibfnamefont {M.~M.}\ \bibnamefont
  {Sheikh-Jabbari}},\ }\bibfield  {title} {\bibinfo {title} {{Solution Phase
  Space and Conserved Charges: A General Formulation for Charges Associated
  with Exact Symmetries}},\ }\href {https://doi.org/10.1103/PhysRevD.93.044074}
  {\bibfield  {journal} {\bibinfo  {journal} {Phys. Rev. D}\ }\textbf {\bibinfo
  {volume} {93}},\ \bibinfo {pages} {044074} (\bibinfo {year} {2016})},\
  \Eprint {https://arxiv.org/abs/1512.05584} {arXiv:1512.05584 [hep-th]}
  \BibitemShut {NoStop}%
\bibitem [{\citenamefont {Hajian}(2016)}]{Hajian:2016kxx}%
  \BibitemOpen
  \bibfield  {author} {\bibinfo {author} {\bibfnamefont {K.}~\bibnamefont
  {Hajian}},\ }\bibfield  {title} {\bibinfo {title} {{Conserved charges and
  first law of thermodynamics for Kerr\textendash{}de Sitter black holes}},\
  }\href {https://doi.org/10.1007/s10714-016-2108-4} {\bibfield  {journal}
  {\bibinfo  {journal} {Gen. Rel. Grav.}\ }\textbf {\bibinfo {volume} {48}},\
  \bibinfo {pages} {114} (\bibinfo {year} {2016})},\ \Eprint
  {https://arxiv.org/abs/1602.05575} {arXiv:1602.05575 [gr-qc]} \BibitemShut
  {NoStop}%
\bibitem [{\citenamefont {Astorino}\ \emph {et~al.}(2016)\citenamefont
  {Astorino}, \citenamefont {Comp\`ere}, \citenamefont {Oliveri},\ and\
  \citenamefont {Vandevoorde}}]{Astorino:2016hls}%
  \BibitemOpen
  \bibfield  {author} {\bibinfo {author} {\bibfnamefont {M.}~\bibnamefont
  {Astorino}}, \bibinfo {author} {\bibfnamefont {G.}~\bibnamefont {Comp\`ere}},
  \bibinfo {author} {\bibfnamefont {R.}~\bibnamefont {Oliveri}},\ and\ \bibinfo
  {author} {\bibfnamefont {N.}~\bibnamefont {Vandevoorde}},\ }\bibfield
  {title} {\bibinfo {title} {{Mass of Kerr-Newman black holes in an external
  magnetic field}},\ }\href {https://doi.org/10.1103/PhysRevD.94.024019}
  {\bibfield  {journal} {\bibinfo  {journal} {Phys. Rev. D}\ }\textbf {\bibinfo
  {volume} {94}},\ \bibinfo {pages} {024019} (\bibinfo {year} {2016})},\
  \Eprint {https://arxiv.org/abs/1602.08110} {arXiv:1602.08110 [gr-qc]}
  \BibitemShut {NoStop}%
\bibitem [{\citenamefont {Gao}\ and\ \citenamefont {Gao}(2023)}]{Gao:2023luj}%
  \BibitemOpen
  \bibfield  {author} {\bibinfo {author} {\bibfnamefont {Y.}~\bibnamefont
  {Gao}}\ and\ \bibinfo {author} {\bibfnamefont {S.}~\bibnamefont {Gao}},\
  }\bibfield  {title} {\bibinfo {title} {{General mass formula for charged
  Kerr-AdS black holes}},\ }\href@noop {} {\  (\bibinfo {year} {2023})},\
  \Eprint {https://arxiv.org/abs/2304.10290} {arXiv:2304.10290 [gr-qc]}
  \BibitemShut {NoStop}%
\bibitem [{\citenamefont {Peng}\ \emph {et~al.}(2021)\citenamefont {Peng},
  \citenamefont {Zou},\ and\ \citenamefont {Liu}}]{Peng:2020cfy}%
  \BibitemOpen
  \bibfield  {author} {\bibinfo {author} {\bibfnamefont {J.-J.}\ \bibnamefont
  {Peng}}, \bibinfo {author} {\bibfnamefont {C.-L.}\ \bibnamefont {Zou}},\ and\
  \bibinfo {author} {\bibfnamefont {H.-F.}\ \bibnamefont {Liu}},\ }\bibfield
  {title} {\bibinfo {title} {{A Komar-like integral for mass and angular
  momentum of asymptotically AdS black holes in Einstein gravity}},\ }\href
  {https://doi.org/10.1088/1402-4896/ac1cd1} {\bibfield  {journal} {\bibinfo
  {journal} {Phys. Scripta}\ }\textbf {\bibinfo {volume} {96}},\ \bibinfo
  {pages} {125207} (\bibinfo {year} {2021})},\ \Eprint
  {https://arxiv.org/abs/2008.06733} {arXiv:2008.06733 [gr-qc]} \BibitemShut
  {NoStop}%
\bibitem [{\citenamefont {Ghosh}\ \emph {et~al.}(2016)\citenamefont {Ghosh},
  \citenamefont {Tannukij},\ and\ \citenamefont {Wongjun}}]{Ghosh:2015cva}%
  \BibitemOpen
  \bibfield  {author} {\bibinfo {author} {\bibfnamefont {S.~G.}\ \bibnamefont
  {Ghosh}}, \bibinfo {author} {\bibfnamefont {L.}~\bibnamefont {Tannukij}},\
  and\ \bibinfo {author} {\bibfnamefont {P.}~\bibnamefont {Wongjun}},\
  }\bibfield  {title} {\bibinfo {title} {{A class of black holes in dRGT
  massive gravity and their thermodynamical properties}},\ }\href
  {https://doi.org/10.1140/epjc/s10052-016-3943-x} {\bibfield  {journal}
  {\bibinfo  {journal} {Eur. Phys. J. C}\ }\textbf {\bibinfo {volume} {76}},\
  \bibinfo {pages} {119} (\bibinfo {year} {2016})},\ \Eprint
  {https://arxiv.org/abs/1506.07119} {arXiv:1506.07119 [gr-qc]} \BibitemShut
  {NoStop}%
\bibitem [{\citenamefont {Boonserm}\ \emph {et~al.}(2018)\citenamefont
  {Boonserm}, \citenamefont {Ngampitipan},\ and\ \citenamefont
  {Wongjun}}]{Boonserm:2017qcq}%
  \BibitemOpen
  \bibfield  {author} {\bibinfo {author} {\bibfnamefont {P.}~\bibnamefont
  {Boonserm}}, \bibinfo {author} {\bibfnamefont {T.}~\bibnamefont
  {Ngampitipan}},\ and\ \bibinfo {author} {\bibfnamefont {P.}~\bibnamefont
  {Wongjun}},\ }\bibfield  {title} {\bibinfo {title} {{Greybody factor for
  black holes in dRGT massive gravity}},\ }\href
  {https://doi.org/10.1140/epjc/s10052-018-5975-x} {\bibfield  {journal}
  {\bibinfo  {journal} {Eur. Phys. J. C}\ }\textbf {\bibinfo {volume} {78}},\
  \bibinfo {pages} {492} (\bibinfo {year} {2018})},\ \Eprint
  {https://arxiv.org/abs/1705.03278} {arXiv:1705.03278 [gr-qc]} \BibitemShut
  {NoStop}%
\bibitem [{\citenamefont {Xu}\ \emph {et~al.}(2015)\citenamefont {Xu},
  \citenamefont {Cao},\ and\ \citenamefont {Hu}}]{Xu:2015rfa}%
  \BibitemOpen
  \bibfield  {author} {\bibinfo {author} {\bibfnamefont {J.}~\bibnamefont
  {Xu}}, \bibinfo {author} {\bibfnamefont {L.-M.}\ \bibnamefont {Cao}},\ and\
  \bibinfo {author} {\bibfnamefont {Y.-P.}\ \bibnamefont {Hu}},\ }\bibfield
  {title} {\bibinfo {title} {{P-V criticality in the extended phase space of
  black holes in massive gravity}},\ }\href
  {https://doi.org/10.1103/PhysRevD.91.124033} {\bibfield  {journal} {\bibinfo
  {journal} {Phys. Rev. D}\ }\textbf {\bibinfo {volume} {91}},\ \bibinfo
  {pages} {124033} (\bibinfo {year} {2015})},\ \Eprint
  {https://arxiv.org/abs/1506.03578} {arXiv:1506.03578 [gr-qc]} \BibitemShut
  {NoStop}%
\bibitem [{\citenamefont {Hendi}\ \emph {et~al.}(2017)\citenamefont {Hendi},
  \citenamefont {Mann}, \citenamefont {Panahiyan},\ and\ \citenamefont
  {Eslam~Panah}}]{Hendi:2017fxp}%
  \BibitemOpen
  \bibfield  {author} {\bibinfo {author} {\bibfnamefont {S.~H.}\ \bibnamefont
  {Hendi}}, \bibinfo {author} {\bibfnamefont {R.~B.}\ \bibnamefont {Mann}},
  \bibinfo {author} {\bibfnamefont {S.}~\bibnamefont {Panahiyan}},\ and\
  \bibinfo {author} {\bibfnamefont {B.}~\bibnamefont {Eslam~Panah}},\
  }\bibfield  {title} {\bibinfo {title} {{Van der Waals like behavior of
  topological AdS black holes in massive gravity}},\ }\href
  {https://doi.org/10.1103/PhysRevD.95.021501} {\bibfield  {journal} {\bibinfo
  {journal} {Phys. Rev. D}\ }\textbf {\bibinfo {volume} {95}},\ \bibinfo
  {pages} {021501} (\bibinfo {year} {2017})},\ \Eprint
  {https://arxiv.org/abs/1702.00432} {arXiv:1702.00432 [gr-qc]} \BibitemShut
  {NoStop}%
\bibitem [{\citenamefont {Dolan}\ \emph {et~al.}(2013)\citenamefont {Dolan},
  \citenamefont {Kastor}, \citenamefont {Kubiznak}, \citenamefont {Mann},\ and\
  \citenamefont {Traschen}}]{Dolan:2013ft}%
  \BibitemOpen
  \bibfield  {author} {\bibinfo {author} {\bibfnamefont {B.~P.}\ \bibnamefont
  {Dolan}}, \bibinfo {author} {\bibfnamefont {D.}~\bibnamefont {Kastor}},
  \bibinfo {author} {\bibfnamefont {D.}~\bibnamefont {Kubiznak}}, \bibinfo
  {author} {\bibfnamefont {R.~B.}\ \bibnamefont {Mann}},\ and\ \bibinfo
  {author} {\bibfnamefont {J.}~\bibnamefont {Traschen}},\ }\bibfield  {title}
  {\bibinfo {title} {{Thermodynamic Volumes and Isoperimetric Inequalities for
  de Sitter Black Holes}},\ }\href {https://doi.org/10.1103/PhysRevD.87.104017}
  {\bibfield  {journal} {\bibinfo  {journal} {Phys. Rev. D}\ }\textbf {\bibinfo
  {volume} {87}},\ \bibinfo {pages} {104017} (\bibinfo {year} {2013})},\
  \Eprint {https://arxiv.org/abs/1301.5926} {arXiv:1301.5926 [hep-th]}
  \BibitemShut {NoStop}%
\bibitem [{\citenamefont {Kubiznak}\ \emph {et~al.}(2017)\citenamefont
  {Kubiznak}, \citenamefont {Mann},\ and\ \citenamefont
  {Teo}}]{Kubiznak:2016qmn}%
  \BibitemOpen
  \bibfield  {author} {\bibinfo {author} {\bibfnamefont {D.}~\bibnamefont
  {Kubiznak}}, \bibinfo {author} {\bibfnamefont {R.~B.}\ \bibnamefont {Mann}},\
  and\ \bibinfo {author} {\bibfnamefont {M.}~\bibnamefont {Teo}},\ }\bibfield
  {title} {\bibinfo {title} {{Black hole chemistry: thermodynamics with
  Lambda}},\ }\href {https://doi.org/10.1088/1361-6382/aa5c69} {\bibfield
  {journal} {\bibinfo  {journal} {Class. Quant. Grav.}\ }\textbf {\bibinfo
  {volume} {34}},\ \bibinfo {pages} {063001} (\bibinfo {year} {2017})},\
  \Eprint {https://arxiv.org/abs/1608.06147} {arXiv:1608.06147 [hep-th]}
  \BibitemShut {NoStop}%
\bibitem [{\citenamefont {Banihashemi}\ \emph {et~al.}(2023)\citenamefont
  {Banihashemi}, \citenamefont {Jacobson}, \citenamefont {Svesko},\ and\
  \citenamefont {Visser}}]{Banihashemi:2022htw}%
  \BibitemOpen
  \bibfield  {author} {\bibinfo {author} {\bibfnamefont {B.}~\bibnamefont
  {Banihashemi}}, \bibinfo {author} {\bibfnamefont {T.}~\bibnamefont
  {Jacobson}}, \bibinfo {author} {\bibfnamefont {A.}~\bibnamefont {Svesko}},\
  and\ \bibinfo {author} {\bibfnamefont {M.}~\bibnamefont {Visser}},\
  }\bibfield  {title} {\bibinfo {title} {{The minus sign in the first law of de
  Sitter horizons}},\ }\href {https://doi.org/10.1007/JHEP01(2023)054}
  {\bibfield  {journal} {\bibinfo  {journal} {JHEP}\ }\textbf {\bibinfo
  {volume} {01}},\ \bibinfo {pages} {054}},\ \Eprint
  {https://arxiv.org/abs/2208.11706} {arXiv:2208.11706 [hep-th]} \BibitemShut
  {NoStop}%
\bibitem [{\citenamefont {Tsallis}(1988)}]{Tsallis:1987eu}%
  \BibitemOpen
  \bibfield  {author} {\bibinfo {author} {\bibfnamefont {C.}~\bibnamefont
  {Tsallis}},\ }\bibfield  {title} {\bibinfo {title} {{Possible Generalization
  of Boltzmann-Gibbs Statistics}},\ }\href {https://doi.org/10.1007/BF01016429}
  {\bibfield  {journal} {\bibinfo  {journal} {J. Statist. Phys.}\ }\textbf
  {\bibinfo {volume} {52}},\ \bibinfo {pages} {479} (\bibinfo {year}
  {1988})}\BibitemShut {NoStop}%
\bibitem [{\citenamefont {{Bir{\'o}}}\ and\ \citenamefont
  {{V{\'a}n}}(2011)}]{Biro2011}%
  \BibitemOpen
  \bibfield  {author} {\bibinfo {author} {\bibfnamefont {T.~S.}\ \bibnamefont
  {{Bir{\'o}}}}\ and\ \bibinfo {author} {\bibfnamefont {P.}~\bibnamefont
  {{V{\'a}n}}},\ }\bibfield  {title} {\bibinfo {title} {{Zeroth law
  compatibility of nonadditive thermodynamics}},\ }\href
  {https://doi.org/10.1103/PhysRevE.83.061147} {\bibfield  {journal} {\bibinfo
  {journal} {\pre}\ }\textbf {\bibinfo {volume} {83}},\ \bibinfo {eid} {061147}
  (\bibinfo {year} {2011})},\ \Eprint {https://arxiv.org/abs/1102.0536}
  {arXiv:1102.0536 [physics.gen-ph]} \BibitemShut {NoStop}%
\bibitem [{\citenamefont {Bir\'o}\ \emph {et~al.}(2018)\citenamefont {Bir\'o},
  \citenamefont {Czinner}, \citenamefont {Iguchi},\ and\ \citenamefont
  {V\'an}}]{Biro:2017flp}%
  \BibitemOpen
  \bibfield  {author} {\bibinfo {author} {\bibfnamefont {T.~S.}\ \bibnamefont
  {Bir\'o}}, \bibinfo {author} {\bibfnamefont {V.~G.}\ \bibnamefont {Czinner}},
  \bibinfo {author} {\bibfnamefont {H.}~\bibnamefont {Iguchi}},\ and\ \bibinfo
  {author} {\bibfnamefont {P.}~\bibnamefont {V\'an}},\ }\bibfield  {title}
  {\bibinfo {title} {{Black hole horizons can hide positive heat capacity}},\
  }\href {https://doi.org/10.1016/j.physletb.2018.05.035} {\bibfield  {journal}
  {\bibinfo  {journal} {Phys. Lett. B}\ }\textbf {\bibinfo {volume} {782}},\
  \bibinfo {pages} {228} (\bibinfo {year} {2018})},\ \Eprint
  {https://arxiv.org/abs/1712.09706} {arXiv:1712.09706 [gr-qc]} \BibitemShut
  {NoStop}%
\bibitem [{\citenamefont {Bir\'o}\ \emph {et~al.}(2020)\citenamefont {Bir\'o},
  \citenamefont {Czinner}, \citenamefont {Iguchi},\ and\ \citenamefont
  {V\'an}}]{Biro:2019rms}%
  \BibitemOpen
  \bibfield  {author} {\bibinfo {author} {\bibfnamefont {T.~S.}\ \bibnamefont
  {Bir\'o}}, \bibinfo {author} {\bibfnamefont {V.~G.}\ \bibnamefont {Czinner}},
  \bibinfo {author} {\bibfnamefont {H.}~\bibnamefont {Iguchi}},\ and\ \bibinfo
  {author} {\bibfnamefont {P.}~\bibnamefont {V\'an}},\ }\bibfield  {title}
  {\bibinfo {title} {{Volume dependent extension of Kerr-Newman black hole
  thermodynamics}},\ }\href {https://doi.org/10.1016/j.physletb.2020.135344}
  {\bibfield  {journal} {\bibinfo  {journal} {Phys. Lett. B}\ }\textbf
  {\bibinfo {volume} {803}},\ \bibinfo {pages} {135344} (\bibinfo {year}
  {2020})},\ \Eprint {https://arxiv.org/abs/1912.04547} {arXiv:1912.04547
  [gr-qc]} \BibitemShut {NoStop}%
\bibitem [{\citenamefont {Tavayef}\ \emph {et~al.}(2018)\citenamefont
  {Tavayef}, \citenamefont {Sheykhi}, \citenamefont {Bamba},\ and\
  \citenamefont {Moradpour}}]{Tavayef:2018xwx}%
  \BibitemOpen
  \bibfield  {author} {\bibinfo {author} {\bibfnamefont {M.}~\bibnamefont
  {Tavayef}}, \bibinfo {author} {\bibfnamefont {A.}~\bibnamefont {Sheykhi}},
  \bibinfo {author} {\bibfnamefont {K.}~\bibnamefont {Bamba}},\ and\ \bibinfo
  {author} {\bibfnamefont {H.}~\bibnamefont {Moradpour}},\ }\bibfield  {title}
  {\bibinfo {title} {{Tsallis Holographic Dark Energy}},\ }\href
  {https://doi.org/10.1016/j.physletb.2018.04.001} {\bibfield  {journal}
  {\bibinfo  {journal} {Phys. Lett. B}\ }\textbf {\bibinfo {volume} {781}},\
  \bibinfo {pages} {195} (\bibinfo {year} {2018})},\ \Eprint
  {https://arxiv.org/abs/1804.02983} {arXiv:1804.02983 [gr-qc]} \BibitemShut
  {NoStop}%
\bibitem [{\citenamefont {Moradpour}\ \emph {et~al.}(2018)\citenamefont
  {Moradpour}, \citenamefont {Moosavi}, \citenamefont {Lobo}, \citenamefont
  {Morais~Gra\c{c}a}, \citenamefont {Jawad},\ and\ \citenamefont
  {Salako}}]{Moradpour:2018ivi}%
  \BibitemOpen
  \bibfield  {author} {\bibinfo {author} {\bibfnamefont {H.}~\bibnamefont
  {Moradpour}}, \bibinfo {author} {\bibfnamefont {S.~A.}\ \bibnamefont
  {Moosavi}}, \bibinfo {author} {\bibfnamefont {I.~P.}\ \bibnamefont {Lobo}},
  \bibinfo {author} {\bibfnamefont {J.~P.}\ \bibnamefont {Morais~Gra\c{c}a}},
  \bibinfo {author} {\bibfnamefont {A.}~\bibnamefont {Jawad}},\ and\ \bibinfo
  {author} {\bibfnamefont {I.~G.}\ \bibnamefont {Salako}},\ }\bibfield  {title}
  {\bibinfo {title} {{Thermodynamic approach to holographic dark energy and the
  R\'enyi entropy}},\ }\href {https://doi.org/10.1140/epjc/s10052-018-6309-8}
  {\bibfield  {journal} {\bibinfo  {journal} {Eur. Phys. J. C}\ }\textbf
  {\bibinfo {volume} {78}},\ \bibinfo {pages} {829} (\bibinfo {year} {2018})},\
  \Eprint {https://arxiv.org/abs/1803.02195} {arXiv:1803.02195
  [physics.gen-ph]} \BibitemShut {NoStop}%
\bibitem [{\citenamefont {Saridakis}(2020)}]{Saridakis:2020zol}%
  \BibitemOpen
  \bibfield  {author} {\bibinfo {author} {\bibfnamefont {E.~N.}\ \bibnamefont
  {Saridakis}},\ }\bibfield  {title} {\bibinfo {title} {{Barrow holographic
  dark energy}},\ }\href {https://doi.org/10.1103/PhysRevD.102.123525}
  {\bibfield  {journal} {\bibinfo  {journal} {Phys. Rev. D}\ }\textbf {\bibinfo
  {volume} {102}},\ \bibinfo {pages} {123525} (\bibinfo {year} {2020})},\
  \Eprint {https://arxiv.org/abs/2005.04115} {arXiv:2005.04115 [gr-qc]}
  \BibitemShut {NoStop}%
\bibitem [{\citenamefont {Sayahian~Jahromi}\ \emph {et~al.}(2018)\citenamefont
  {Sayahian~Jahromi}, \citenamefont {Moosavi}, \citenamefont {Moradpour},
  \citenamefont {Morais~Gra\c{c}a}, \citenamefont {Lobo}, \citenamefont
  {Salako},\ and\ \citenamefont {Jawad}}]{SayahianJahromi:2018irq}%
  \BibitemOpen
  \bibfield  {author} {\bibinfo {author} {\bibfnamefont {A.}~\bibnamefont
  {Sayahian~Jahromi}}, \bibinfo {author} {\bibfnamefont {S.~A.}\ \bibnamefont
  {Moosavi}}, \bibinfo {author} {\bibfnamefont {H.}~\bibnamefont {Moradpour}},
  \bibinfo {author} {\bibfnamefont {J.~P.}\ \bibnamefont {Morais~Gra\c{c}a}},
  \bibinfo {author} {\bibfnamefont {I.~P.}\ \bibnamefont {Lobo}}, \bibinfo
  {author} {\bibfnamefont {I.~G.}\ \bibnamefont {Salako}},\ and\ \bibinfo
  {author} {\bibfnamefont {A.}~\bibnamefont {Jawad}},\ }\bibfield  {title}
  {\bibinfo {title} {{Generalized entropy formalism and a new holographic dark
  energy model}},\ }\href {https://doi.org/10.1016/j.physletb.2018.02.052}
  {\bibfield  {journal} {\bibinfo  {journal} {Phys. Lett. B}\ }\textbf
  {\bibinfo {volume} {780}},\ \bibinfo {pages} {21} (\bibinfo {year} {2018})},\
  \Eprint {https://arxiv.org/abs/1802.07722} {arXiv:1802.07722 [gr-qc]}
  \BibitemShut {NoStop}%
\bibitem [{\citenamefont {Nojiri}\ \emph
  {et~al.}(2022{\natexlab{b}})\citenamefont {Nojiri}, \citenamefont
  {Odintsov},\ and\ \citenamefont {Faraoni}}]{Nojiri:2022aof}%
  \BibitemOpen
  \bibfield  {author} {\bibinfo {author} {\bibfnamefont {S.}~\bibnamefont
  {Nojiri}}, \bibinfo {author} {\bibfnamefont {S.~D.}\ \bibnamefont
  {Odintsov}},\ and\ \bibinfo {author} {\bibfnamefont {V.}~\bibnamefont
  {Faraoni}},\ }\bibfield  {title} {\bibinfo {title} {{From nonextensive
  statistics and black hole entropy to the holographic dark universe}},\ }\href
  {https://doi.org/10.1103/PhysRevD.105.044042} {\bibfield  {journal} {\bibinfo
   {journal} {Phys. Rev. D}\ }\textbf {\bibinfo {volume} {105}},\ \bibinfo
  {pages} {044042} (\bibinfo {year} {2022}{\natexlab{b}})},\ \Eprint
  {https://arxiv.org/abs/2201.02424} {arXiv:2201.02424 [gr-qc]} \BibitemShut
  {NoStop}%
\bibitem [{\citenamefont {Nojiri}\ \emph
  {et~al.}(2022{\natexlab{c}})\citenamefont {Nojiri}, \citenamefont
  {Odintsov},\ and\ \citenamefont {Paul}}]{Nojiri:2022dkr}%
  \BibitemOpen
  \bibfield  {author} {\bibinfo {author} {\bibfnamefont {S.}~\bibnamefont
  {Nojiri}}, \bibinfo {author} {\bibfnamefont {S.~D.}\ \bibnamefont
  {Odintsov}},\ and\ \bibinfo {author} {\bibfnamefont {T.}~\bibnamefont
  {Paul}},\ }\bibfield  {title} {\bibinfo {title} {{Early and late universe
  holographic cosmology from a new generalized entropy}},\ }\href
  {https://doi.org/10.1016/j.physletb.2022.137189} {\bibfield  {journal}
  {\bibinfo  {journal} {Phys. Lett. B}\ }\textbf {\bibinfo {volume} {831}},\
  \bibinfo {pages} {137189} (\bibinfo {year} {2022}{\natexlab{c}})},\ \Eprint
  {https://arxiv.org/abs/2205.08876} {arXiv:2205.08876 [gr-qc]} \BibitemShut
  {NoStop}%
\bibitem [{\citenamefont {Nojiri}\ \emph
  {et~al.}(2022{\natexlab{d}})\citenamefont {Nojiri}, \citenamefont
  {Odintsov},\ and\ \citenamefont {Faraoni}}]{Nojiri:2022ljp}%
  \BibitemOpen
  \bibfield  {author} {\bibinfo {author} {\bibfnamefont {S.}~\bibnamefont
  {Nojiri}}, \bibinfo {author} {\bibfnamefont {S.~D.}\ \bibnamefont
  {Odintsov}},\ and\ \bibinfo {author} {\bibfnamefont {V.}~\bibnamefont
  {Faraoni}},\ }\bibfield  {title} {\bibinfo {title} {{New Entropies, Black
  Holes, and Holographic Dark Energy}},\ }\href
  {https://doi.org/10.1007/s10511-023-09759-1} {\bibfield  {journal} {\bibinfo
  {journal} {Astrophysics}\ }\textbf {\bibinfo {volume} {65}},\ \bibinfo
  {pages} {534} (\bibinfo {year} {2022}{\natexlab{d}})},\ \Eprint
  {https://arxiv.org/abs/2208.10235} {arXiv:2208.10235 [gr-qc]} \BibitemShut
  {NoStop}%
\bibitem [{\citenamefont {Nojiri}\ \emph
  {et~al.}(2022{\natexlab{e}})\citenamefont {Nojiri}, \citenamefont
  {Odintsov},\ and\ \citenamefont {Paul}}]{Nojiri:2022nmu}%
  \BibitemOpen
  \bibfield  {author} {\bibinfo {author} {\bibfnamefont {S.}~\bibnamefont
  {Nojiri}}, \bibinfo {author} {\bibfnamefont {S.~D.}\ \bibnamefont
  {Odintsov}},\ and\ \bibinfo {author} {\bibfnamefont {T.}~\bibnamefont
  {Paul}},\ }\bibfield  {title} {\bibinfo {title} {{Modified cosmology from the
  thermodynamics of apparent horizon}},\ }\href
  {https://doi.org/10.1016/j.physletb.2022.137553} {\bibfield  {journal}
  {\bibinfo  {journal} {Phys. Lett. B}\ }\textbf {\bibinfo {volume} {835}},\
  \bibinfo {pages} {137553} (\bibinfo {year} {2022}{\natexlab{e}})},\ \Eprint
  {https://arxiv.org/abs/2211.02822} {arXiv:2211.02822 [gr-qc]} \BibitemShut
  {NoStop}%
\bibitem [{\citenamefont {Cohen}\ \emph {et~al.}(1999)\citenamefont {Cohen},
  \citenamefont {Kaplan},\ and\ \citenamefont {Nelson}}]{Cohen:1998zx}%
  \BibitemOpen
  \bibfield  {author} {\bibinfo {author} {\bibfnamefont {A.~G.}\ \bibnamefont
  {Cohen}}, \bibinfo {author} {\bibfnamefont {D.~B.}\ \bibnamefont {Kaplan}},\
  and\ \bibinfo {author} {\bibfnamefont {A.~E.}\ \bibnamefont {Nelson}},\
  }\bibfield  {title} {\bibinfo {title} {{Effective field theory, black holes,
  and the cosmological constant}},\ }\href
  {https://doi.org/10.1103/PhysRevLett.82.4971} {\bibfield  {journal} {\bibinfo
   {journal} {Phys. Rev. Lett.}\ }\textbf {\bibinfo {volume} {82}},\ \bibinfo
  {pages} {4971} (\bibinfo {year} {1999})},\ \Eprint
  {https://arxiv.org/abs/hep-th/9803132} {arXiv:hep-th/9803132} \BibitemShut
  {NoStop}%
\bibitem [{\citenamefont {Nakarachinda}\ \emph {et~al.}(2022)\citenamefont
  {Nakarachinda}, \citenamefont {Pongkitivanichkul}, \citenamefont {Samart},
  \citenamefont {Tannukij},\ and\ \citenamefont
  {Wongjun}}]{Nakarachinda:2022mlz}%
  \BibitemOpen
  \bibfield  {author} {\bibinfo {author} {\bibfnamefont {R.}~\bibnamefont
  {Nakarachinda}}, \bibinfo {author} {\bibfnamefont {C.}~\bibnamefont
  {Pongkitivanichkul}}, \bibinfo {author} {\bibfnamefont {D.}~\bibnamefont
  {Samart}}, \bibinfo {author} {\bibfnamefont {L.}~\bibnamefont {Tannukij}},\
  and\ \bibinfo {author} {\bibfnamefont {P.}~\bibnamefont {Wongjun}},\
  }\bibfield  {title} {\bibinfo {title} {{Holographic dark energy from the
  anti\textendash{}de Sitter black hole}},\ }\href
  {https://doi.org/10.1103/PhysRevD.105.123524} {\bibfield  {journal} {\bibinfo
   {journal} {Phys. Rev. D}\ }\textbf {\bibinfo {volume} {105}},\ \bibinfo
  {pages} {123524} (\bibinfo {year} {2022})},\ \Eprint
  {https://arxiv.org/abs/2201.09715} {arXiv:2201.09715 [gr-qc]} \BibitemShut
  {NoStop}%
\bibitem [{\citenamefont {Hawking}(1976)}]{Hawking:1976ra}%
  \BibitemOpen
  \bibfield  {author} {\bibinfo {author} {\bibfnamefont {S.~W.}\ \bibnamefont
  {Hawking}},\ }\bibfield  {title} {\bibinfo {title} {{Breakdown of
  Predictability in Gravitational Collapse}},\ }\href
  {https://doi.org/10.1103/PhysRevD.14.2460} {\bibfield  {journal} {\bibinfo
  {journal} {Phys. Rev. D}\ }\textbf {\bibinfo {volume} {14}},\ \bibinfo
  {pages} {2460} (\bibinfo {year} {1976})}\BibitemShut {NoStop}%
\end{thebibliography}%

\end{document}